\documentclass[10pt,twocolumn,letterpaper]{article}

\usepackage{iccv}
\usepackage{times}
\usepackage{epsfig}
\usepackage{graphicx}
\usepackage{amsmath}
\usepackage{amssymb}
\usepackage{bm}

\usepackage{wrapfig}
\usepackage{subcaption}

\usepackage[ruled]{algorithm2e} 
\usepackage{booktabs}
\usepackage{tabularx}

\usepackage[breaklinks=true,bookmarks=false]{hyperref}

\iccvfinalcopy 

\def\iccvPaperID{****} 

\ificcvfinal\fi

\begin{document}

\title{Clusterplot: High-dimensional Cluster Visualization}

\author{Or Malkai
\\
Tel-Aviv University\\
{\tt\small ormalkai@gmail.com}
\and
Min Lu \\
Shenzhen University\\
{\tt\small lumin.vis@gmail.com}
\and
Daniel Cohen-Or \\
Tel-Aviv University\\
{\tt\small cohenor@gmail.com}
}

\maketitle

\ificcvfinal\thispagestyle{empty}\fi

\begin{abstract}
We present \textit{Clusterplot}, a multi-class high-dimensional data visualization tool designed to visualize cluster-level information offering an intuitive understanding of the cluster inter-relations. 
Our unique plots leverage 2D blobs devised to convey the geometrical and topological characteristics of clusters within the high-dimensional data, and their pairwise relations, such that general inter-cluster behavior is easily interpretable in the plot. Class identity supervision is utilized to drive the measuring of relations among clusters in high-dimension, particularly, \textit{proximity} and \textit{overlap}, which are then 
reflected spatially through the 2D blobs. We demonstrate the strength of our clusterplots and their ability to deliver a clear and intuitive informative exploration experience for high-dimensional clusters characterized by complex structure and significant overlap.
\end{abstract}


\section{Introduction}

High-dimensional data exploration is inherently challenging; intractability of visualization in the original dimension necessarily calls for dimensionality reduction, shifting the focus to the difficulties associated with this type of operation. 
Scatterplotting the data on a 2D canvas is a common visual means achieved by the application of some form of reduction, but projecting to lower dimensional space often causes visual clutter, compromising the ease of visual exploration.
Many techniques have been suggested to alleviate this clutter, separating the data into clusters by maintaining similarity between the low- and high-dimensional distributions \cite{engel2012survey}.
These techniques are \textit{unsupervised}; they do not assume any annotation of data, thus the groundtruth division into clusters is unknown.

In this work, conversely, we target multi-class high-dimensional data visualization with a known classification.
Exploration of multi-class data that is largely high-dimensional has become the focus of many, with the recent growing popularity of the deep learning paradigm, a substantial part of which is tagged (e.g., MNIST or ImageNet).
Moreover, many of these tasks incorporate a form of latent space embedding \cite{KhoslaYaoJayadevaprakashFeiFei_FGVC2011}\cite{imagenet_cvpr09}, 
where large datasets are mapped into a high-dimensional space as part of, e.g., classification and retrieval efforts. As such, exploration of the data in its embedded state is often crucial for the evaluation and appraisal of the employed model, but the high dimensionality poses a challenge. 

\begin{figure}[t]
\begin{center}
  \begin{subfigure}[t]{0.48\columnwidth}
    \includegraphics[width=\linewidth]{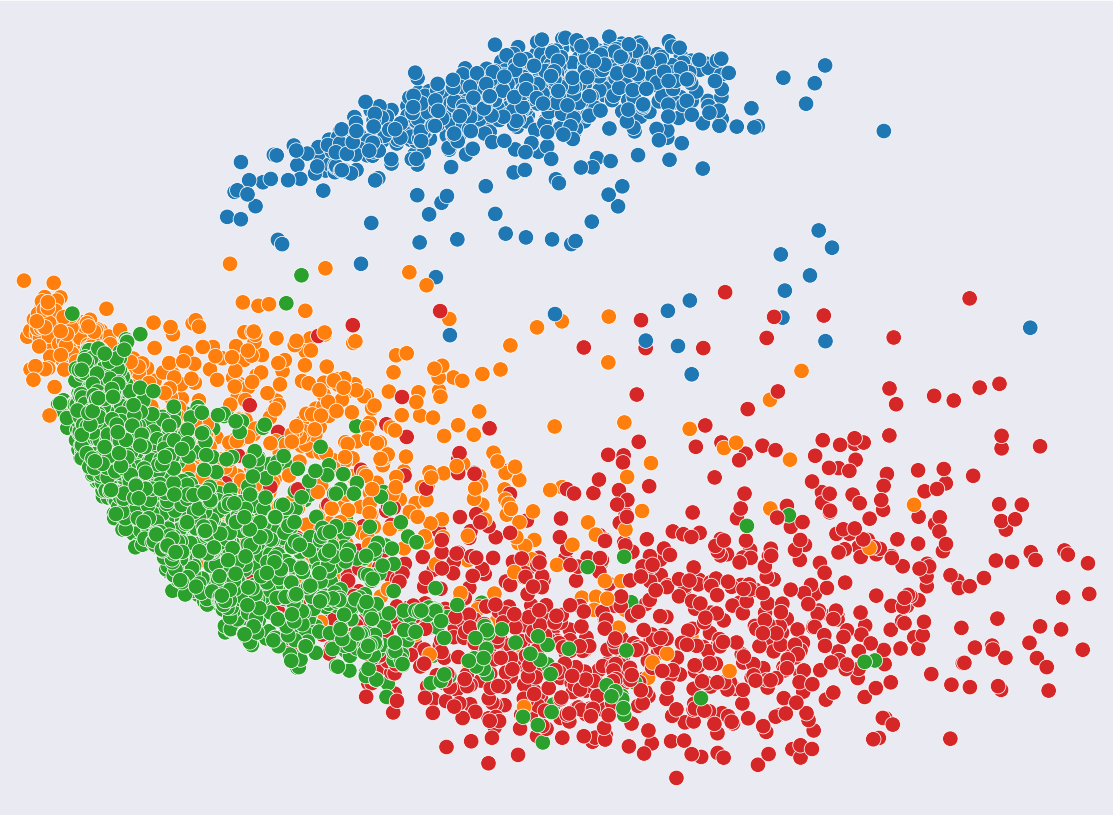}
    \caption{PCA}
  \end{subfigure}
  \begin{subfigure}[t]{0.48\columnwidth}
    \includegraphics[width=\linewidth]{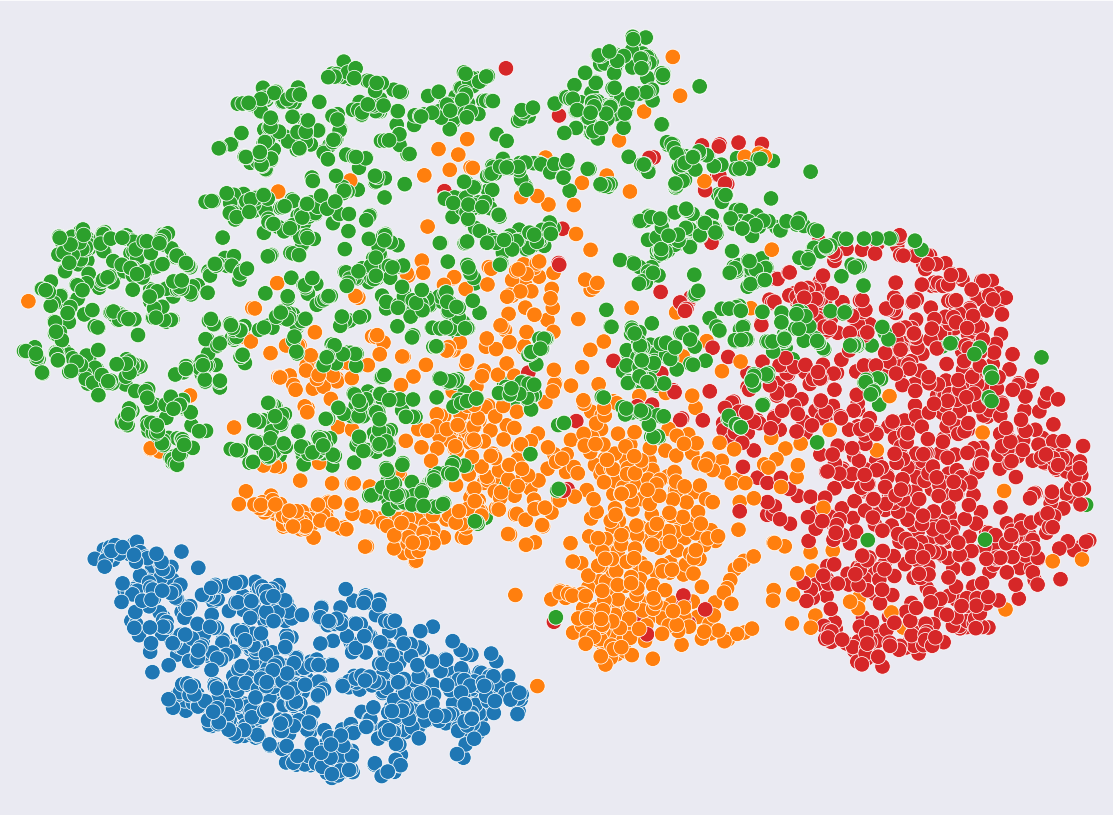}
    \caption{t-SNE}
  \end{subfigure}
  \begin{subfigure}[t]{0.48\columnwidth}
    \includegraphics[width=\linewidth]{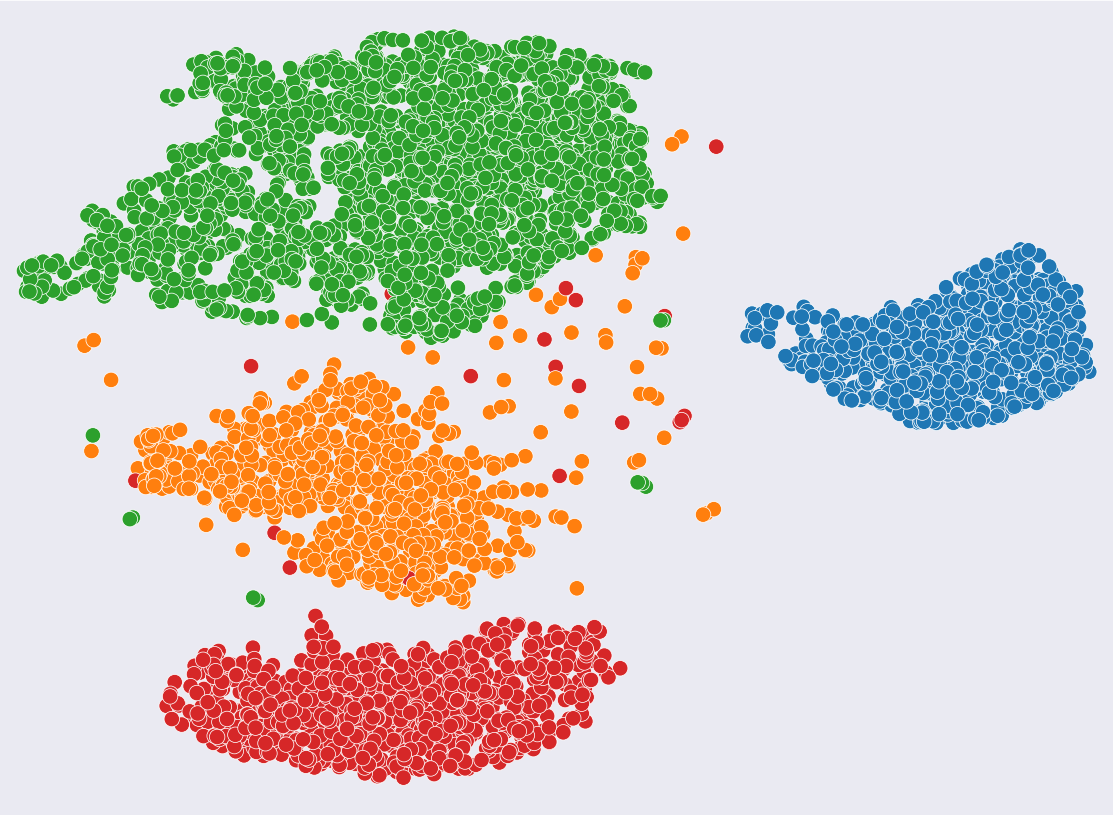}
    \caption{Supervised UMAP}
  \end{subfigure}
  \begin{subfigure}[t]{0.48\columnwidth}
    \includegraphics[width=\linewidth]{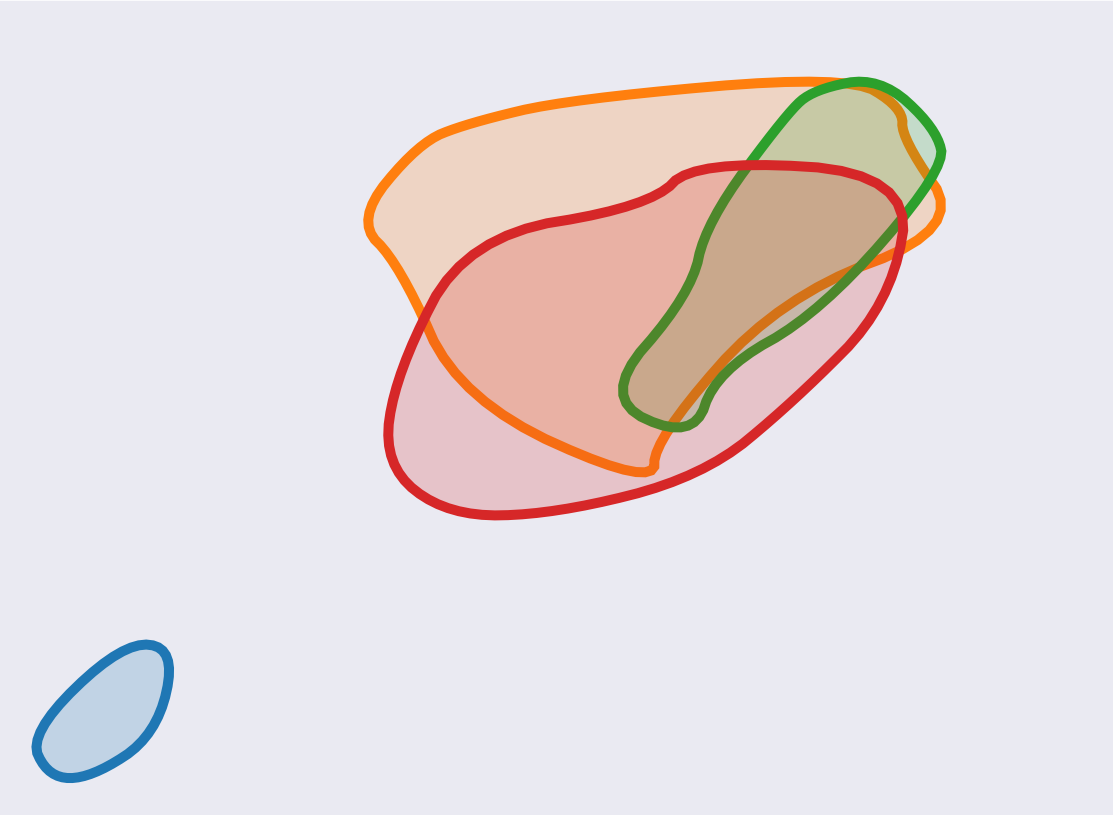}
    \caption{Clusterplot}
  \end{subfigure}
  \caption{Visualization of PCA, t-SNE, Supervised UMAP and Clusterplot of four clusters. Note the significant differences of plots representing the same high-dimensional data. }
  \label{fig:4Fashion}
\end{center}
\end{figure}

For example, Figure \ref{fig:4Fashion} 
shows different plots generated by different techniques of a relatively simple  
dataset with four clusters. As can be seen, even with just four clusters, common visualization techniques may yield rather different plots displaying the same high-dimensional data. Figure \ref{fig:4Fashion}(a)(b) shows the PCA and t-SNE projections, which are unsupervised and unclear what quantitative information of clusters these plots convey. In (c), the result of supervised UMAP is shown as reference, presenting a completely different view of the data, as it aims to display the clusters apart from each other. 
Figure \ref{fig:4Fashion}(d) 
presents the result of such a \textit{Clusterplot}, under supervision of class labels, where clusters are displayed as colored blobs, and their pairwise spatial relations reflect their quantitative inter-cluster relations (i.e., proximity and overlap among clusters) as measured in high-dimensional space.
Note that in contrast to UMAP, the clusterplot aims to preserve and display the overlaps among the clusters.

Clearly, as the number of clusters increases, so does plot clutter, raising legitimate concerns pertaining to the aptitude of these 2D plots as exploration tools for high-dimensional data, and motivating our cluster-driven plotting approach as a strong alternative, where the plots reflect quantitative measures.

Our premise is that clusters of high-dimensional entities are often geometrically and topologically complex, and the portrayal of their minute details in a 2D plot in a way that is interpretable and comprehensive to the viewer, is essentially unattainable. The clusterplots presented in this paper do not presume to illustrate each and every detail, and are, rather, simple and abstract by design, aiming to reflect proximity and overlap among clusters, to provide important spatial cues that can be highly indicative of the nature and characteristics of the embedded data. 

With the data being classified, the \textit{proximity} and \textit{overlap} between pairs of classes can be quantitatively measured. Clusterplots are driven and optimized towards this high level of data abstraction, by analyzing the cluster's density and applying dimensionality reduction on representative points only.

\begin{figure}[t]
\begin{center}
  \begin{subfigure}[t]{0.90\columnwidth}
    \includegraphics[width=\linewidth]{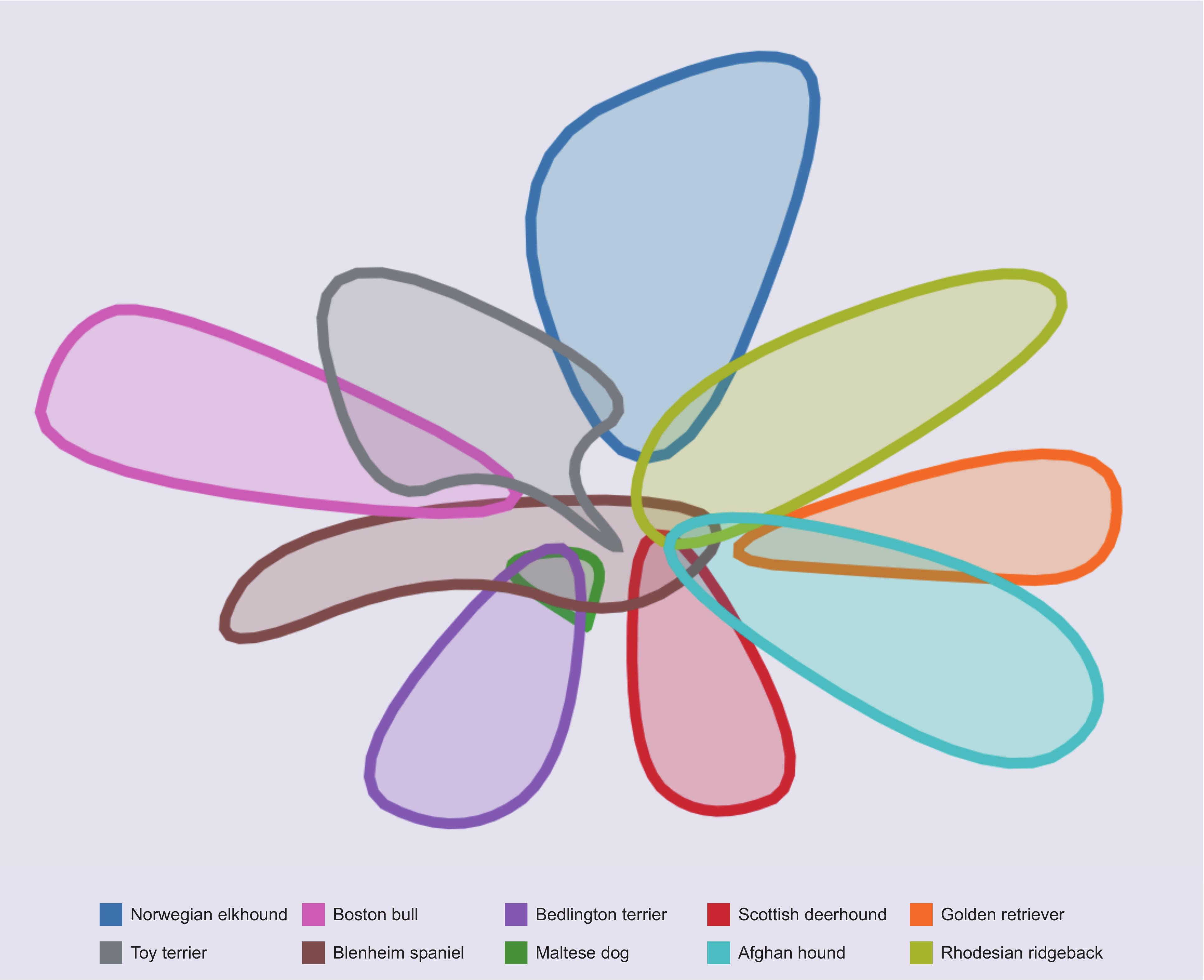}
    \caption{Clusterplot}
  \end{subfigure}
  \begin{subfigure}[t]{0.45\columnwidth}
    \includegraphics[width=\linewidth]{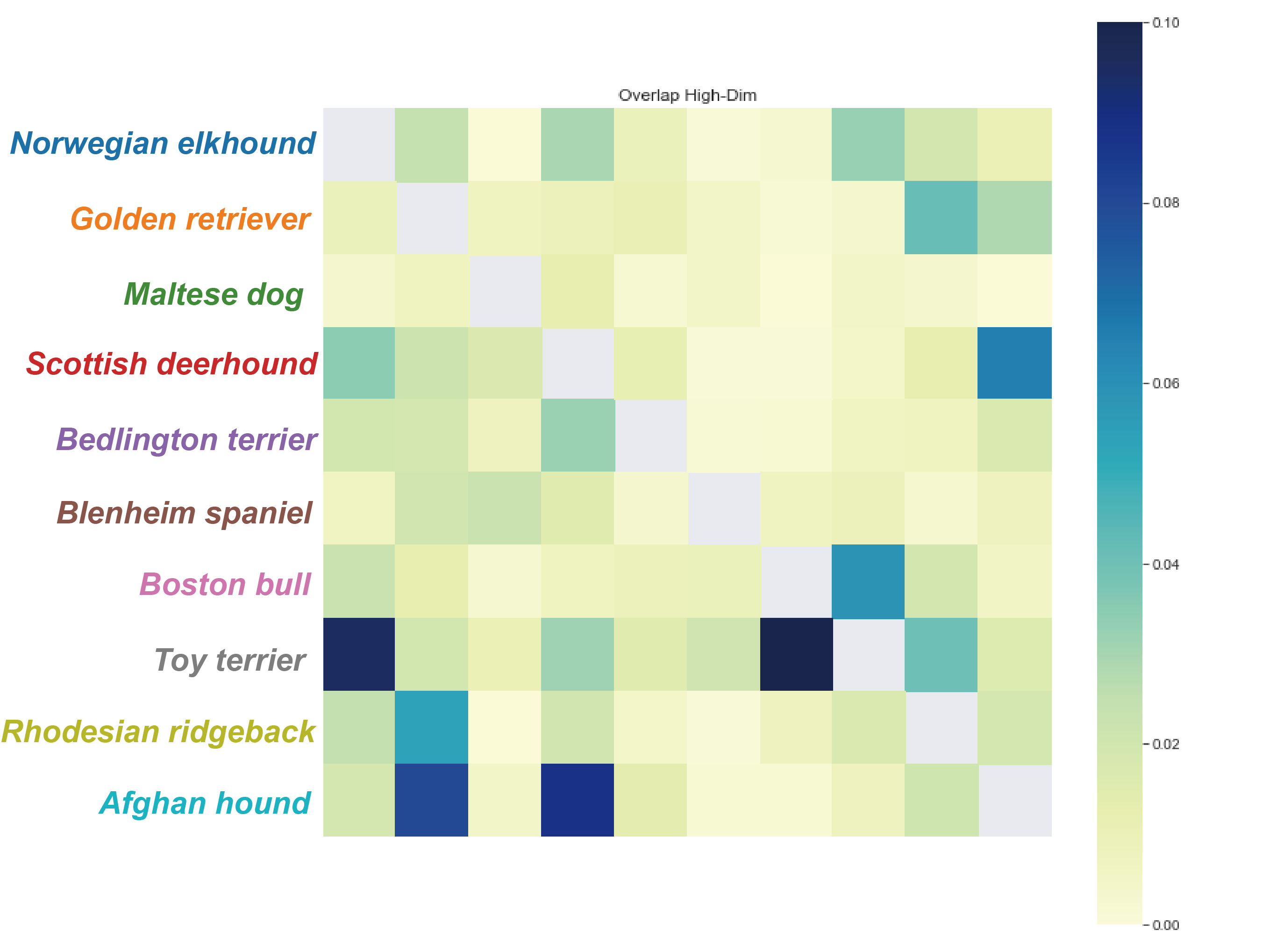}
    \caption{Overlap}
  \end{subfigure}
  \begin{subfigure}[t]{0.45\columnwidth}
    \includegraphics[width=\linewidth]{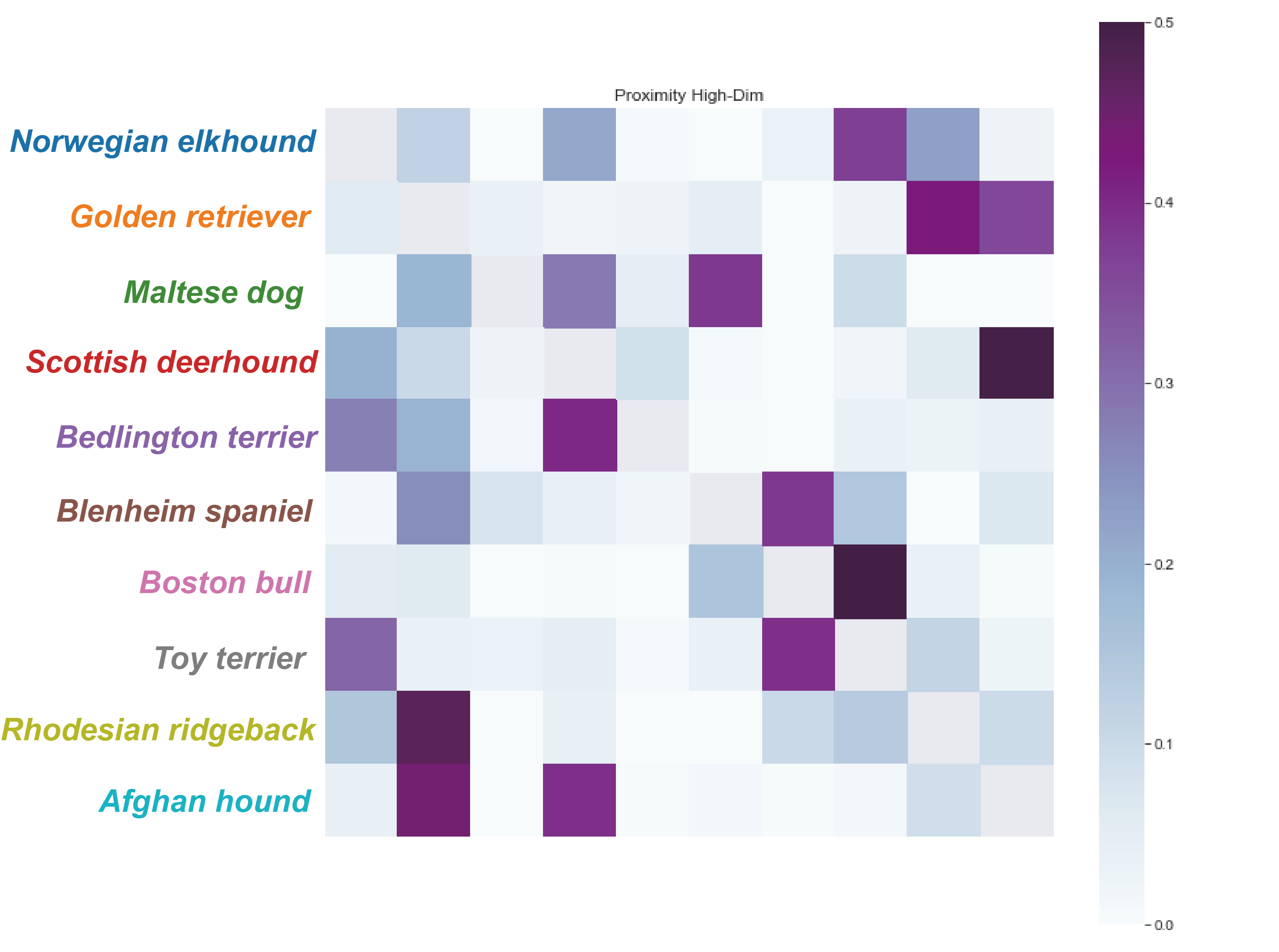}
    \caption{Proximity}
  \end{subfigure}
  \caption{Clusterplot, overlap and proximity matrices measured on VGG features (4096 dimension) on ten classes of Stanford Dogs Dataset \cite{KhoslaYaoJayadevaprakashFeiFei_FGVC2011}. The spatial relations in the clusterplot aim to reflect the overlap and proximity measures which are presented in the matrices.}
  \label{fig:teaser}
\end{center}
\end{figure}

We demonstrate the strength and capacity of our clusterplots to alleviate efforts to explore and visualize high-dimensional data, on a variety of examples. Specifically, we focus on classified datasets of images encoded by deep neural networks, and show that a cluster plot provides a quick intuitive visual means to reflect upon the performance of the employed model, thereby assisting users in their quest to evaluate their work, and make educated decisions moving forward. A controlled user study was conducted to examine the efficiency of clusterplots in perceiving the class inter-relations. The results show that \textit{Clusterplot} form a prominent and more intuitive visual representation of the spatial relation among the clusters over the baseline design in the form of \textit{Matrices} (see Figure \ref{fig:teaser}).

\section{Related Work}
\label{sec:related_work}


In this section, we present an overview of cluster visualization techniques, and then discuss research progress in high dimensional data visualization, and more related, dimensionality reduction techniques. 

\subsection{Clusters Visualization}

There are three major approaches to cluster visualizations. The first is to assign a new visual channel, or better allocate visual cues of the original visualization, to encode the information of clustering. Scatterplot is the most conventional technique using points for high dimensional data visualization. A large body of research focuses on enhancing the perception of the different classes in scatterplots.
Wang et al.~\cite{wang2018optimizing} optimize coloring strategies to better perceptually separate the different classes.
Chen et al.~\cite{chen2018using} suggest per-class flickering to improve the interpretation of multiple classes in Scatterplot Matrices with overlapping points.
For better showing the joined relation among point clouds, Luboschik et al.~\cite{LuboschikRadloffSchumann2010} present a color weaving technique by defining the color interlacing pattern in overlap regions. 
Chen et al.~\cite{chen2014visual} propose a multi-class down sampling method to alleviate the overdrawing and make intersecting features more visible. 

The second approach is to visually enclosure objects in a cluster with contours to explicitly pronounce classes.
For example, Collins et al.~\cite{collins2009bubble} propose \textit{Bubble sets} to delineate set membership among points.  
Splatterplot~\cite{MayorgaGleicher2013} draw boundary around the main distribution of a class, while retaining the visibility of outliers. 
Jo et al.~\cite{jo2019declarative} summarize the design space of aggregated multiclass maps and present a declarative grammar to construct the aggregation. 
Lu et al.~\cite{lu2019winglets} design local strokes, \textit{Winglets}, to shape the global structure of classes using Gestalt Closure Principle. 

Different from those works which enhance the perception of clusters of a given embedding, the third approach to cluster visualization is to visually abstract each cluster and focus on expressing the relationship among clusters. 
Euler diagram~\cite{euler2003lettres} developed by Leonard Euler in the 18 centuries is one of the first ideas along this direction. Euler diagrams represent clusters (or sets) as shapes and encode their relationship via overlapping among them. 
Venn diagram~\cite{venn1880diagrammatic} is a specific form of Euler diagram, which draws all possible logical relations between sets. 
Due to the restrictive regulation, the difficulty in Venn diagram comprehension scales up when the number of sets increases~\cite{ruskey1997survey}. %
Many follow-up works refine the design of Euler and Venn diagrams, but generalized for a small number of sets only. For example, Micallef et al. ~\cite{micallef2014eulerape} proposed drawing area-proportional Venn diagram of three sets only.
On the other hand, some works derive new visual forms from the conventional Euler or Venn diagrams, by relaxing the regulation on continuous and convexity of the shapes. 
Simonetto et al.~\cite{simonetto2009fully} model intersections between sets as a graph where nodes represent intersected regions between sets, and construct contours to bound these regions of sets in the graph. 
Riche et al.~\cite{riche2010untangling} present Compact Rectangular Euler Diagrams, which simplify the shapes of sets to compact rectangles connected by links. 
%

Our work continues with the same spirit as Euler diagrams, manipulating 2D shapes to show the cluster relationships in the high dimensional data. Different from those techniques with explicit logical information as input (e.g., the sets of labels), our method takes the high dimensional data, which is implicit, and represents the relationship in high dimensional space by 2D blobs in low dimensional space. 



\subsection{High-dimensional Data Visualization}



%

Visualization of high-dimensional data has been extensively studied in the field of Information Visualization~\cite{keim2002information}. Over the years, many techniques with rich visual expressiveness have been developed~\cite{liu2016visualizing}. One large group is to visually project each data item from high dimensional space into 2D or 3D space. For example, Scatterplot is widely used to draw high-dimensional data as points in Cartesian coordinates after dimensionality reduction~\cite{maaten2008visualizing}. SPLOM (Scatterplot Matrix) organizes bivariate scatterplots into a matrix to support the observation of multiple bivariate relationship simultaneously~\cite{wilkinson2006high}. GSPLOM is its variation when taking nominal attributes into account~\cite{im2013gplom}. 

Different from representing data as points, Parallel Coordinates draw each data as a polyline across parallel dimension axes~\cite{inselberg1985plane}. Star plot (or Radar chart) arranges axes in a radial layout, for a more compact layout~\cite{kandogan2000star}. To handle large amount of data, pixel-based visualization is proposed to encode data values to individual pixels and generate separate display for each dimension~\cite{keim1994visdb}. On the other hand, some works explore the representation of data item by an iconic figure (i.e., glyph), such as Chernoff face~\cite{chernoff1973use} using different facial features to encode attributes.  Different from accessing to
the details of each data point, our method abstracts from those details and aims to visualize the relation among the classes, a meta representation of the high-dimensional data. 

Some works also visualize the meta information of high dimensional data, but with different focus. For example, Yang et al.~\cite{yang2003interactive} study dimensions by similarity and visualize them in a radial hierarchical layout, for better dimension management in high-dimensional analysis. To abstract the distribution, skeleton-based principal graph is constructed for scatterplots to summarize local density and deviations~\cite{reddy2008generating}, or to indicate the shape and orientation of point  distribution~\cite{matute2017skeleton}. Tapping to human capability of understanding natural terrains, Weber et al.~\cite{weber2007topological} construct 2D landscape to visualize the contour tree of high-dimensional functions. Similarly, Oesterling et al.~\cite{oesterling2011visualization} abstract multidimensional point clouds by extracting the join tree from density estimation and present it as a topological landscape. Engel et al.~\cite{engel2011structural} design structural decomposition tree to show the hierarchy in high dimensional data while keeping details of value coordinates visible. In this work, we focus on the visualization of clusters spatial relation.


\subsection{Dimensionality Reduction}

Over the years, a large number reduction techniques has been introduced for multidimensional data \cite{engel2012survey}. The vast majority of the methods are unsupervised, where the data is not tagged. These unsupervised methods help analyzing the high-dimensional data and often aim to help clustering it. Principle Component Analysis (PCA) is the most common method used for dimensionality reduction. PCA projects the data on successive orthogonal components such that the maximum variance of the data is preserved \cite{pearson1901liii}. Kernel PCA is an extension of PCA that achieves non-linear dimensionality reduction by using kernels \cite{scholkopf1998nonlinear}. 
Another popular method is Multi-dimensional Scaling (MDS), which searches for low-dimensional embedding such that the distances in the low-dimension reflect the distances in the high dimension \cite{kruskal1964multidimensional}. Isometric Mapping (Isomap) can be seen as an extension of MDS or Kernal PCA, where similarly to MDS, it searches for low-dimensional embedding such that the geodesic distances reflects their corresponding ones in the high-dimension \cite{tenenbaum2000global}. 

In recent years, t-distributed Stochastic Neighbor Embedding (t-SNE) \cite{maaten2008visualizing} gained a lot of popularity. t-SNE is a non-linear method which converts the affinities of the data points to probabilities and uses gradient descent to optimize the KL-divergence between the joint probabilities in the high-dimension and in the low-dimension. Recently, Uniform Manifold Approximation and Projection (UMAP) \cite{mcinnes2018umap} is emerging as an improvement to t-SNE. UMAP uses stochastic gradient descent to optimize a cross-entropy loss, which allows it to better preserve global structures of the data. In our method, we use UMAP as the backbone for embedding representative points of the clusters and defining the initial blobs embedding.

Linear Discriminant Analysis (LDA) is a well-known supervised linear method, which projects the data on the directions that maximizes the classes separation. There are other supervised methods or meta-methods that augment basic unsupervised methods with label information. Bair et al. \cite{bair2006prediction} present a supervised version of PCA, which first chooses a subset of features based on their association with the label. Cheng et al.\cite{cheng2012supervised}  introduce a supervised version of Isomap using pairwise constraints on geodesic distances between data points in the same class or in different classes. Hajderanj et al. \cite{hajderanj2019new} introduce a supervised version of t-SNE with dissimilarity measures related to class information. Similarly, there is a supervised version of UMAP which encourages the different clusters to be embedded separately in the 2D space. Unlike these supervised methods, our method does not aim at using the labels to separate the clusters, but to better visualize the true proximity and overlapping that exist in the data, see Figure \ref{fig:4Fashion}(c-d) which demonstrates that.

\section{Clusterplot}

\subsection{Overview}

The premise of our work is that an exact high-dimensional geometric and topological interface between the clusters is not only impossible to plot in 2D, but also too complex to interpret and comprehend. The clusterplots that we present are designed to be visually easy to understand and yet expressive enough to deliver the inter-relations between the clusters. In particular, \textit{Clusterplot} encodes two inter-cluster relations: proximity and overlap. The key idea is to generate 2D plots that visually approximate these two inter-cluster relations to those in the original high-dimensional space.

In our plots, the clusters are represented by blobs, which is a set of smooth and closed 2D shapes that represent the regions where the corresponding data is embedded (see Figure~\ref{fig:teaser}). Such blobs are an abstraction of what otherwise is commonly represented as scatterplots. Our design choice of using blobs rather than scatterplots assumes that blobs are better visual means to perceive the inter-cluster proximity and overlapping relations. 
We argue that the exact individual data point locations in the embedded space are insignificant in delivering the inter-cluster relations, and contribute to redundant visual clutter. Moreover, since our clusterplot focuses on overlapping, the blobs are displayed with notable thick outlines that allow perceiving and identifying the entire clusters including the overlapping regions.

\begin{figure}[t]
  \centering
    \includegraphics[width=0.49\textwidth]{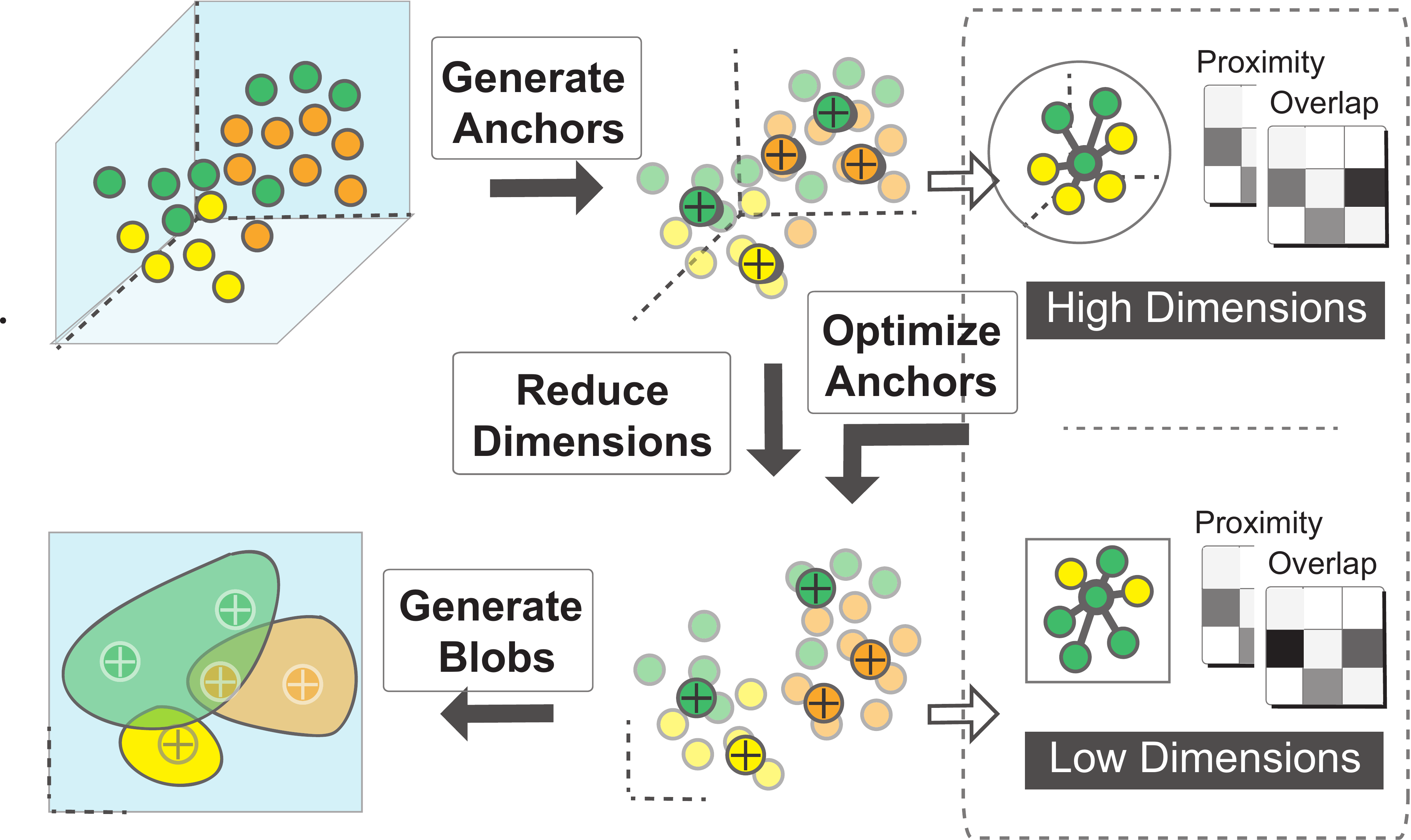}
    \caption{Pipeline of generating \textit{Clusterplot}: a set of anchors is generated for each cluster, projected to 2D space and then iteratively manipulated to minimize the difference between inter-cluster relations in low and high dimensional space.}
  \label{img:overview}
\end{figure}

\begin{figure}[!htb]
  \begin{subfigure}[t]{0.49\columnwidth}
    \includegraphics[width=\linewidth]{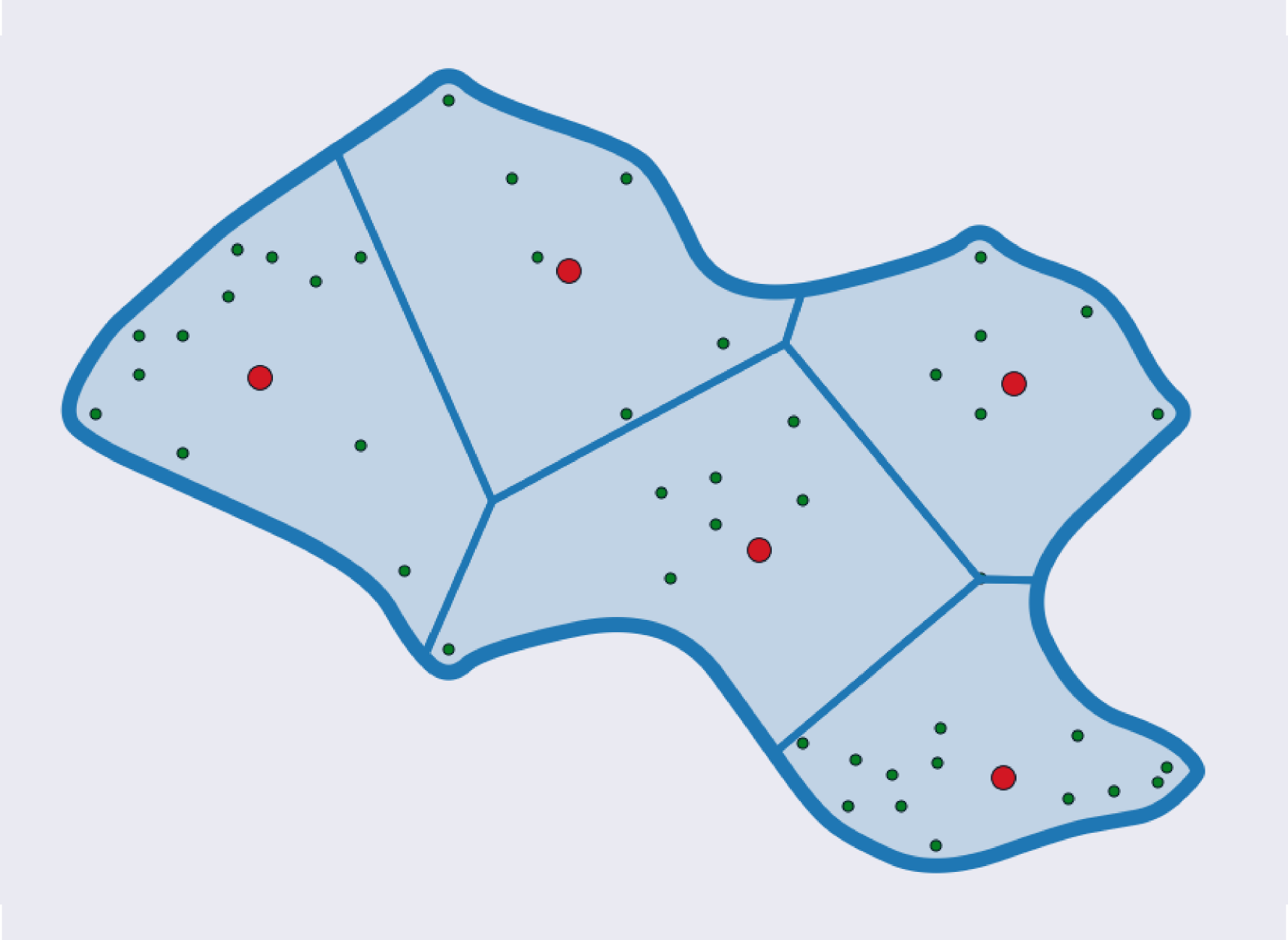}
    \caption{Anchors computed by BIRCH}
  \end{subfigure}
  \begin{subfigure}[t]{0.49\columnwidth}
    \includegraphics[width=\linewidth]{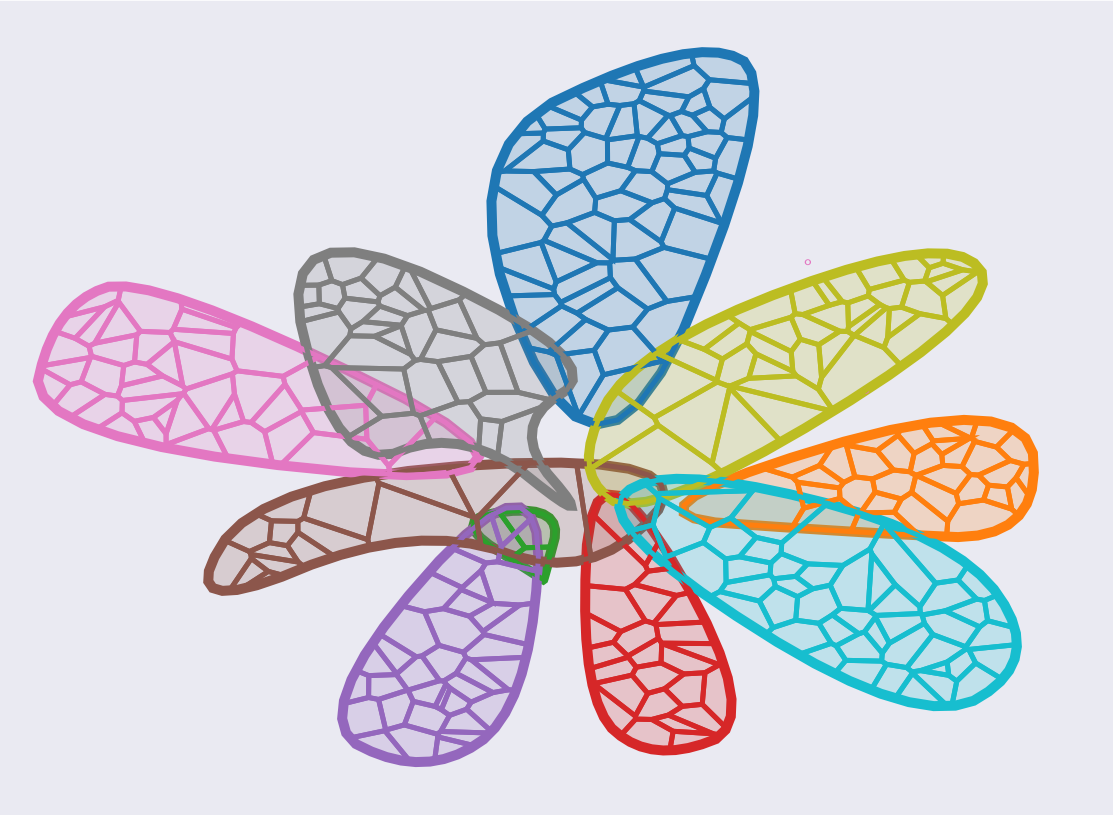}
    \caption{Voronoi form}
  \end{subfigure}
  
  \caption{An example of a cluster with non-uniform density. The green points are the data points, the red points are the anchors, that are the centroids of the sub-clusters, computed by BIRCH. The blob representing the cluster is subdivided into the Voronoi cell of the anchors.} 
  \label{img:birch}
\end{figure}

\textit{Clusterplot} is a result of a cluster-level analysis, yet, the geometry of a blob that represents a cluster is critical for delivering the global inter-cluster relations.  Figure~\ref{img:overview} illustrates the framework and high-level steps in generating clusterplots. To control the blob shapes, we approximate each cluster with a number of sub-clusters, where each sub-cluster has a representative point that we refer to as an \textit{anchor} (see the red points in Figure \ref{img:birch}(a)). These sub-cluster anchors are an intermediate representation, which is not displayed to the user, but only used to compute the blob's shape to reflect the inter-relation among the clusters. The proximity and overlapping values are measured based on the anchors and integrated to define the inter-cluster relations. These two measures are defined using K-nearest neighbors (KNN) means without using any dimension-dependent metric, hence allowing to compare inter-cluster relations in low and high dimensions, and in particular, between the input clusters in high-dimension and their counterparts in the 2D. Anchors are initially projected to lower dimensional space (2D in our case) and then iteratively optimized towards the minimal difference between inter-cluster relations in low and high dimensional space. In the end, simplified blobs are generated around each set of 2D anchors.

\subsection{Density-agnostic representation}

Let us consider the classification problem of $N$ data points $X = \{x_i: 1 \leq i \leq N\}$ to a set of labels $L=\{1,...,M\}$. Let 
 $Y = \{y_i: 1 \leq i \leq N, y_i \in L\}$, then $$D=\{(x_i,y_i): 1 \leq i \leq N, x_i\in X,y_i \in Y\}.$$

We define a sub-clustering function by $S:D \to \mathbb{N}$, then sub-cluster $j$ is defined as $$SC_j=\{x_i:S(x_i,y_i )=j\}.$$ 
An intuitive way to define such sub-clustering function $S$ is to apply a clustering algorithm for each label separately. Then the representative points, or \textit{anchors}, can be defined as the centroids of these sub-clusters $$A_j = \frac{\sum_{x_i \in SC_j}{x_i}}{|SC_j|},$$ yielding anchors for each label, separately.

Any clustering algorithm, such as K-means or hierarchical clustering can be used. However a critical design choice here is that the sub-clustering and the anchors definition assumes that the clusters have a non-uniform density. See the partition in Figure \ref{img:birch}(a), where the cells do not contain similar number of points. That is, the sub-clusters do not represent a similar number of data points, but similar area. As we shall elaborate below, we use the Balanced Iterative Reducing and Clustering using Hierarchies (BIRCH) algorithm \cite{zhang1996birch} to create sub-clusters that represent a similar area of the given cluster, and by taking a representative point per sub-cluster it creates a uniform density of the cluster.

The BIRCH can be seen as a data reduction method since it reduces the data to a set of sub-clusters. The BIRCH algorithm has two parameters: a threshold $T$ and branching factor $B$, where the latter is not essential for our needs. The threshold parameter determines the radius of the sub-clusters, and all sub-clusters computed by BIRCH must satisfy the requirement that their radius is less than $T$. Therefore, the threshold parameter allows reducing the data to a set of sub-clusters with maximum radius $T$, yielding a representation of the data with a uniform density. As $T$ gets larger, there are less sub-clusters and vice-versa. The anchors are defined as the centroids of the sub-clusters. See Figure \ref{img:birch}(a) for an example of sub-clusters defined by BIRCH for a cluster with non-uniform density. The anchors are density-agnostic, and form uniform sampling of the cluster area. This allows the blobs representing the cluster shapes, and hence their spatial (proximity) relations. Figure \ref{img:birch}(b) shows the Voronoi diagrams of the anchors.


\subsection{Measuring Inter-cluster relations}

Once the anchors are defined, we can reduce their dimension using some dimensionality-reduction algorithm, such as PCA, MDS,  or t-SNE. 
We use state-of-the-art UMAP algorithm (with \textit{min-dist} set to $1$), to embed the anchors to 2D. 
Each anchor $A_j$ in the high-dimension is represented by $LA_j$ in the 2D embedded space.

Note that reducing the dimension of the anchors only, allows preserving the local (the shape of cluster) and global geometry (the inter-cluster relations) of the clusters, while significantly simplifying the spatial optimization. 
Our clusterplots encode two inter-cluster
relations: proximity and overlapping. We define these two measures using K-nearest neighbors (KNN) means, without using any dimension-dependent metrics.
\\\\
\textbf{Overlap.}
The overlap inter-cluster relation is an $N_A \times N_A$ matrix, where $N_A$ is the number of anchors. Each cell $MA^O_{i,j}$ is the overlap between the regions represented by anchor $i$ and anchor $j$. The overlap is defined as the ratio between edges from sub-cluster $i$ to sub-cluster $j$ in the $directed$ $K_O$-Nearest Neighbor Graph on the data points:
$$EA_{i,j} = \{(p_1,p_2) : p_1 \in SC_i, p_2 \in SC_j \}$$
then:
$$MA^O_{i,j} = \frac{|EA_{i,j}|}{\sum_l |EA_{i,l}|}.$$

To get an overlap measure in cluster-level, we define an $M \times M$ matrix, where $M$ is the number of labels, such that:
$$EL_{i,j} = \{(p_1,p_2) : p_1 \in Label_i, p_2 \in Label_j \},$$
and then,$$ML^O_{i,j} = \frac{|EL_{i,j}|}{\sum_l |EL_{i,l}|}.$$ 

\textbf{Proximity.}
The inter-cluster proximity relation is encoded into an $M \times M$ matrix. The proximity between clusters $Label_i$ and $Label_j$, denoted by $ML^P_{i,j}$, is defined as the ratio between edges from $Label_i$ to $Label_j$ in the $directed$ $K_P$-Nearest Neighbor Graph (KNNG) on the anchors, with the constraint that there are no edges between anchors of the same cluster:
$$EL_{i,j} = \{(p_1,p_2) : p_1 \in Label_i, p_2 \in Label_j \}.$$
Then:
$$ML^P_{i,j} = \frac{|EL_{i,j}|}{\sum_l |EL_{i,l}|}$$


\textbf{Inter-Cluster Relations}

\begin{figure}[!htb]
  \begin{subfigure}[t]{0.49\columnwidth}
    \includegraphics[width=\linewidth]{img/VGGAnimals_Overlap_High-Dim_iter_0.png}
    \caption{Overlap High-Dim}
  \end{subfigure}
  \begin{subfigure}[t]{0.49\columnwidth}
    \includegraphics[width=\linewidth]{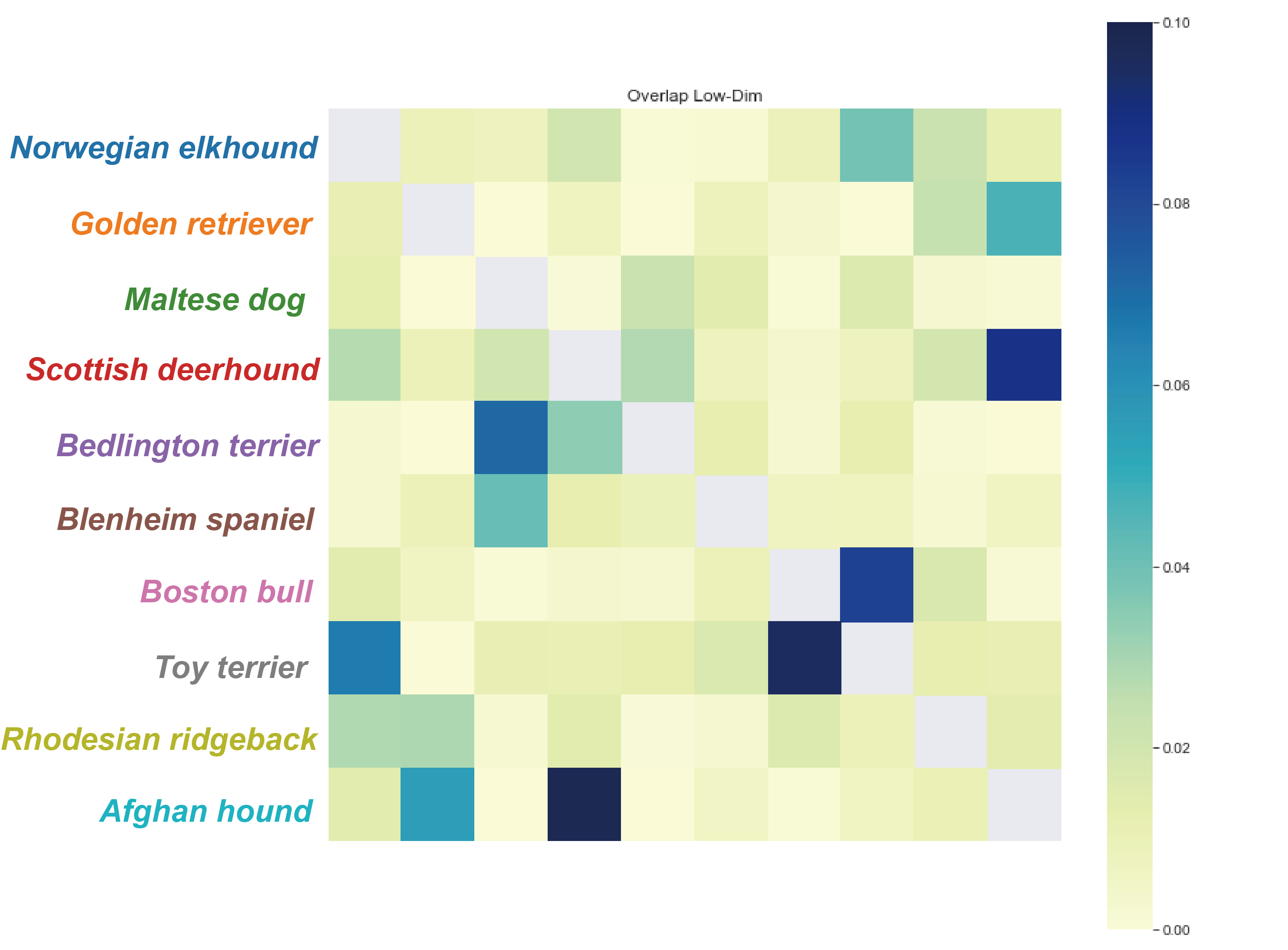}
    \caption{Overlap Low-Dim}
  \end{subfigure}
  
    \begin{subfigure}[t]{0.49\columnwidth}
    \includegraphics[width=\linewidth]{img/VGGAnimals_Proximity_High-Dim_iter_0.png}
    \caption{Proximity High-Dim}
  \end{subfigure}
  \begin{subfigure}[t]{0.49\columnwidth}
    \includegraphics[width=\linewidth]{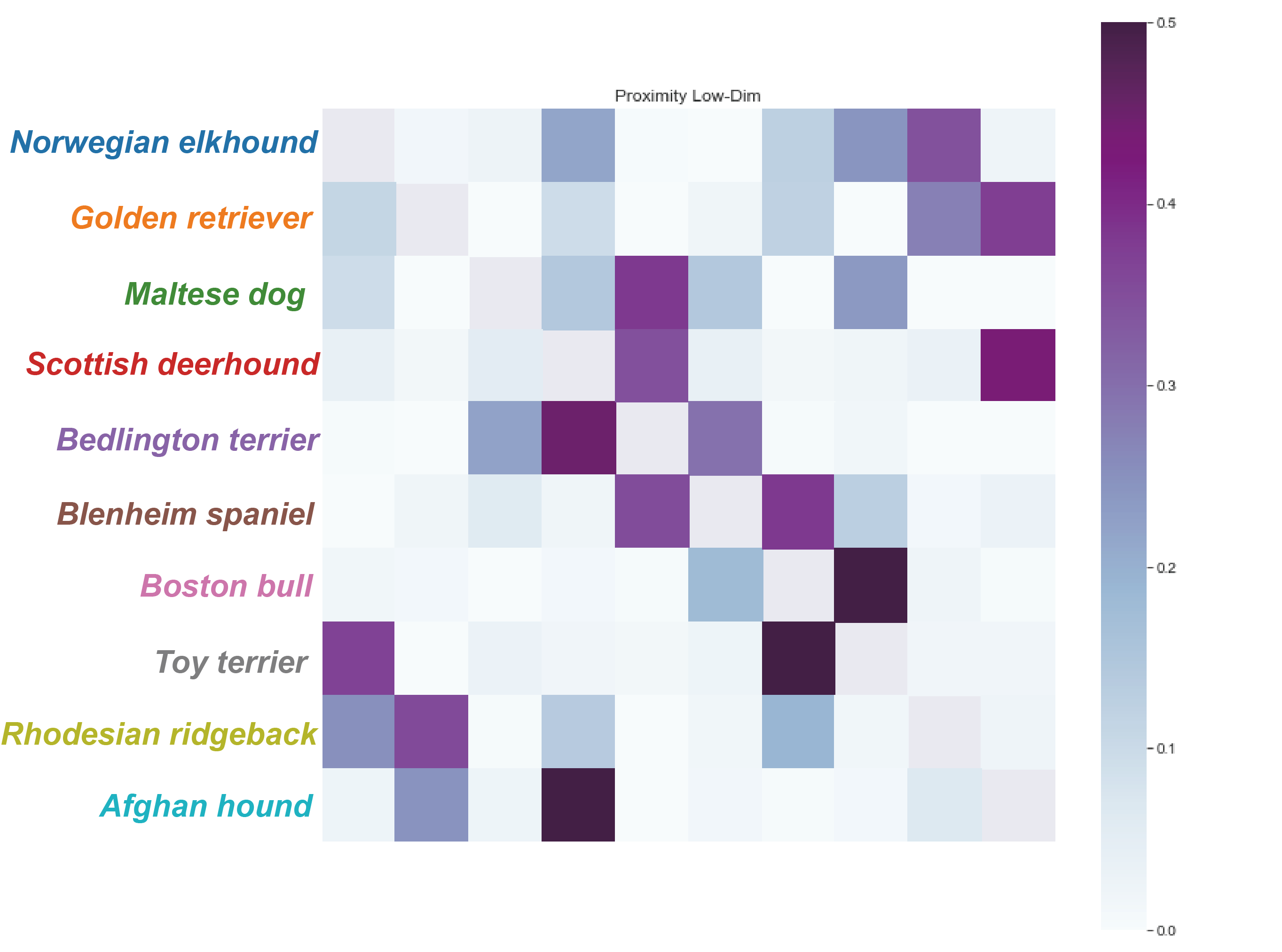}
    \caption{Proximity Low-Dim}
  \end{subfigure}
  \caption{Overlap and proximity matrices in high and low dimension. }
  \label{fig:HighVsLowMatrices}
\end{figure}

Given $K_O$ and $K_P$ of the K-Nearest Neighbor Graphs, the calculation of the overlap and proximity matrices in the high dimension is straightforward based on the description above. However, the calculation of the overlap and proximity matrices in the low dimension is not straightforward, since we reduced the dimension of the sparse set of anchors only. Therefore, we first generate  virtual data points in the low dimension as follows: First, for each cluster, we find a close bounding shape of its anchors using the alpha shape (default $\alpha=1$) algorithm, while ignoring outlier using the Local Outlier Factor (LOF) filter \cite{breunig2000lof}. Then, we partition the bounding shape into Voronoi regions induced by the anchors, such that every Voronoi region in 2D  corresponds with a sub-cluster in the high dimension. We spread virtual data points in each Voronoi region as the number of data points in the corresponded sub-cluster in the high-dimension. These virtual points are used to measure the overlap among the blobs in 2D. Figure \ref{fig:HighVsLowMatrices} shows the measured overlap and proximity matrices in both high-dimension and 2D on Stanford Dogs
Dataset.

\subsection{Plot Optimization}

To express the inter-cluster relations in the high-dimension in a 2D plot, we define a loss function between the high-dimensional overlap matrix and the 2D overlap matrix. We use the Maximum Absolute Error as the loss metric. Denote by $HMA^O$ and $LMA^O$ the high-dimensional and low-dimensional overlap matrices, respectively, then:
$$MAE = max_{i,j}\{|HMA^O_{i,j} - LMA^O_{i,j}|\}.$$
Clearly, the optimization problem is non-linear and non-convex, therefore we use a greedy algorithm to minimize the $MAE$. The algorithm picks the sub-clusters with the maximum difference of overlap between the high-dimension and the low-dimension, then if these sub-clusters are too close to each other the algorithm pushes sub-cluster $i$ from sub-cluster $j$ on the line connecting anchor $i$ and anchor $j$, and if the sub-clusters are too far apart it pulls sub-cluster $i$ closer to sub-cluster $j$ on the line connecting anchor $i$ and anchor $j$.

\begin{algorithm}[h!]
 \KwIn{$I$: number of iterations, $l$ : Learning rate, $\delta$: Stop 
 criteria}
 \For{$iter \in \{1,...,I\}$}
 {
  $MAE \leftarrow max_{i,j}\{|HMA^O_{i,j} - LMA^O_{i,j}|\}$
  
  \eIf{$MAE \le \delta$}
  {
   $STOP$
  }
  {
   $i,j \leftarrow argmax_{i,j}\{|HMA^O_{i,j} - LMA^O_{i,j}|\}$
   
   $\vec{d} \leftarrow \vec{LA_i} - \vec{LA_j}$
   
   \eIf{$HMA^O_{i,j} \le LMA^O_{i,j}$}
  {
    $\vec{LA_i} \leftarrow \vec{LA_i} + l \cdot \vec{d}$
  }
  {
    $\vec{LA_i} \leftarrow \vec{LA_i} - l \cdot \vec{d}$
  }
  }
 }
 \caption{Optimization Algorithm}
\end{algorithm}

\begin{figure}[!htb]
\begin{center}
  \begin{subfigure}[t]{0.49\columnwidth}
    \includegraphics[width=\linewidth]{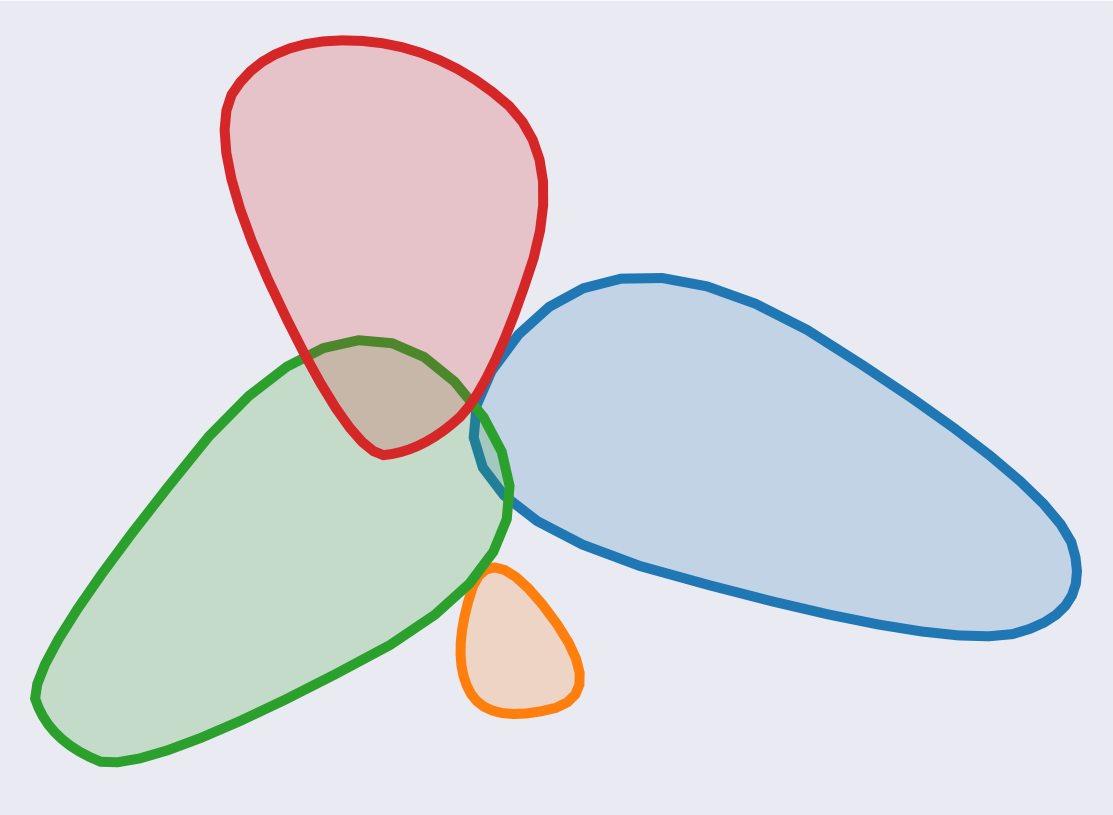}
  \end{subfigure}
  \begin{subfigure}[t]{0.49\columnwidth}
    \includegraphics[width=\linewidth]{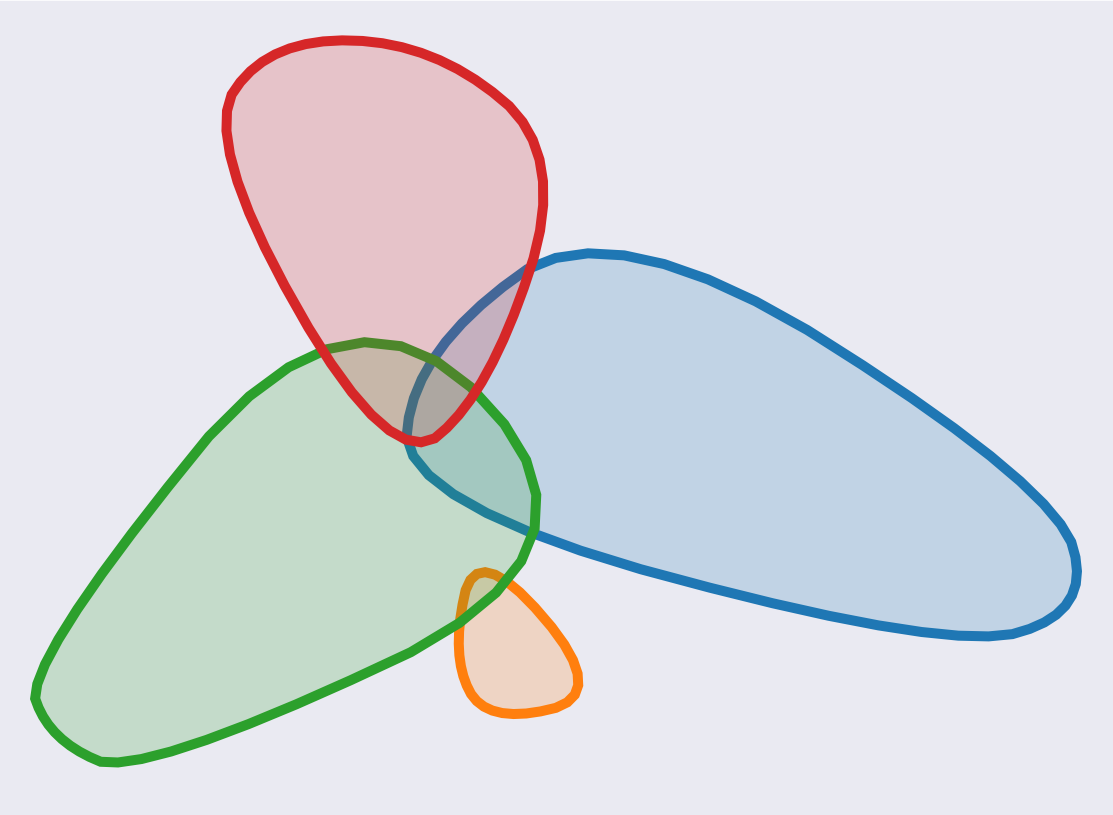}
  \end{subfigure}
  
  \begin{subfigure}[t]{0.49\columnwidth}
    \includegraphics[width=\linewidth]{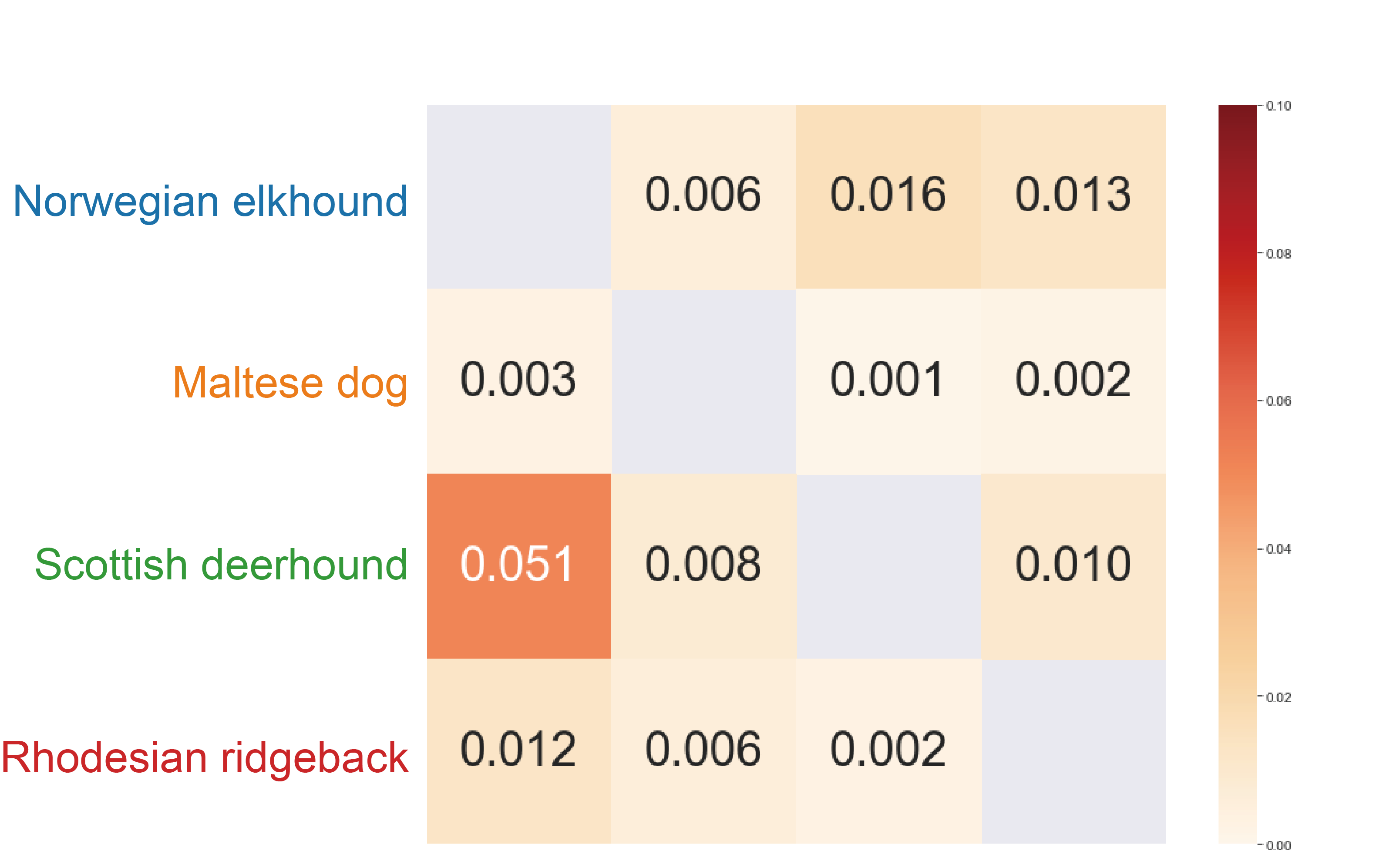}
  \end{subfigure}
  \begin{subfigure}[t]{0.49\columnwidth}
    \includegraphics[width=\linewidth]{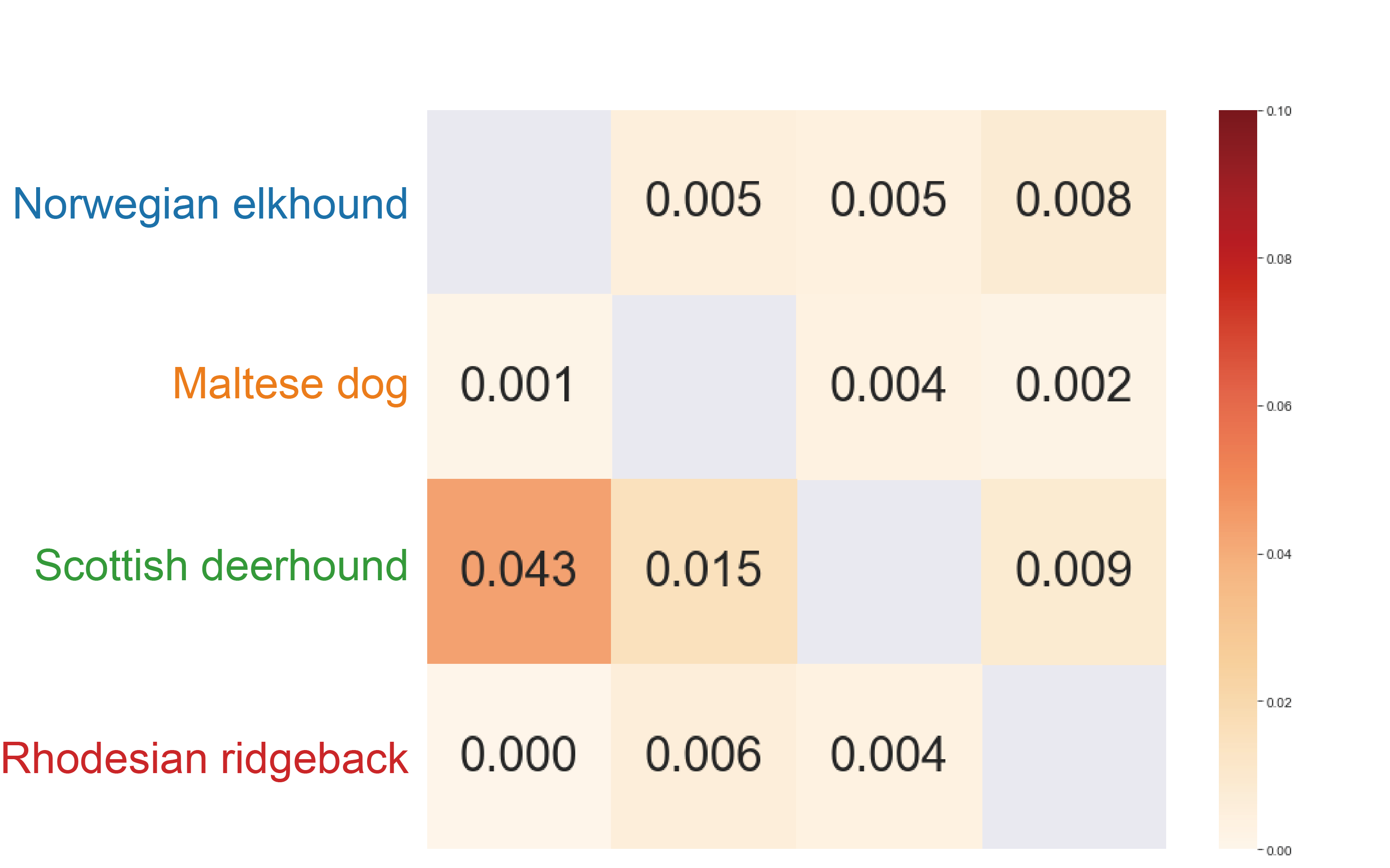}
  \end{subfigure}
  
  \caption{\textit{Clusterplot} and difference matrix between high-dim and 2D before (on the left) and after (on the right) optimization on extracted features by VGG network on images from Stanford Dogs Dataset. 
 }
  \label{fig:OverlapDiffBeforeAndAfter}
  \end{center}
\end{figure}

Figure \ref{fig:OverlapDiffBeforeAndAfter} shows an example of the clusterplot along with the matrix of the differences between the overlap in the high-dimension to the overlap in 2D, before and after the optimization. It can be seen that before the optimization there is little overlap between Scottish Deerhound (green) and the Norwegian Elhkound (blue), but the optimization increases the overlap, which is then closer to the measured overlap in the high-dimension.

\section{Examples}

\begin{figure*}[t]
  \centering
    \includegraphics[width=1.\linewidth]{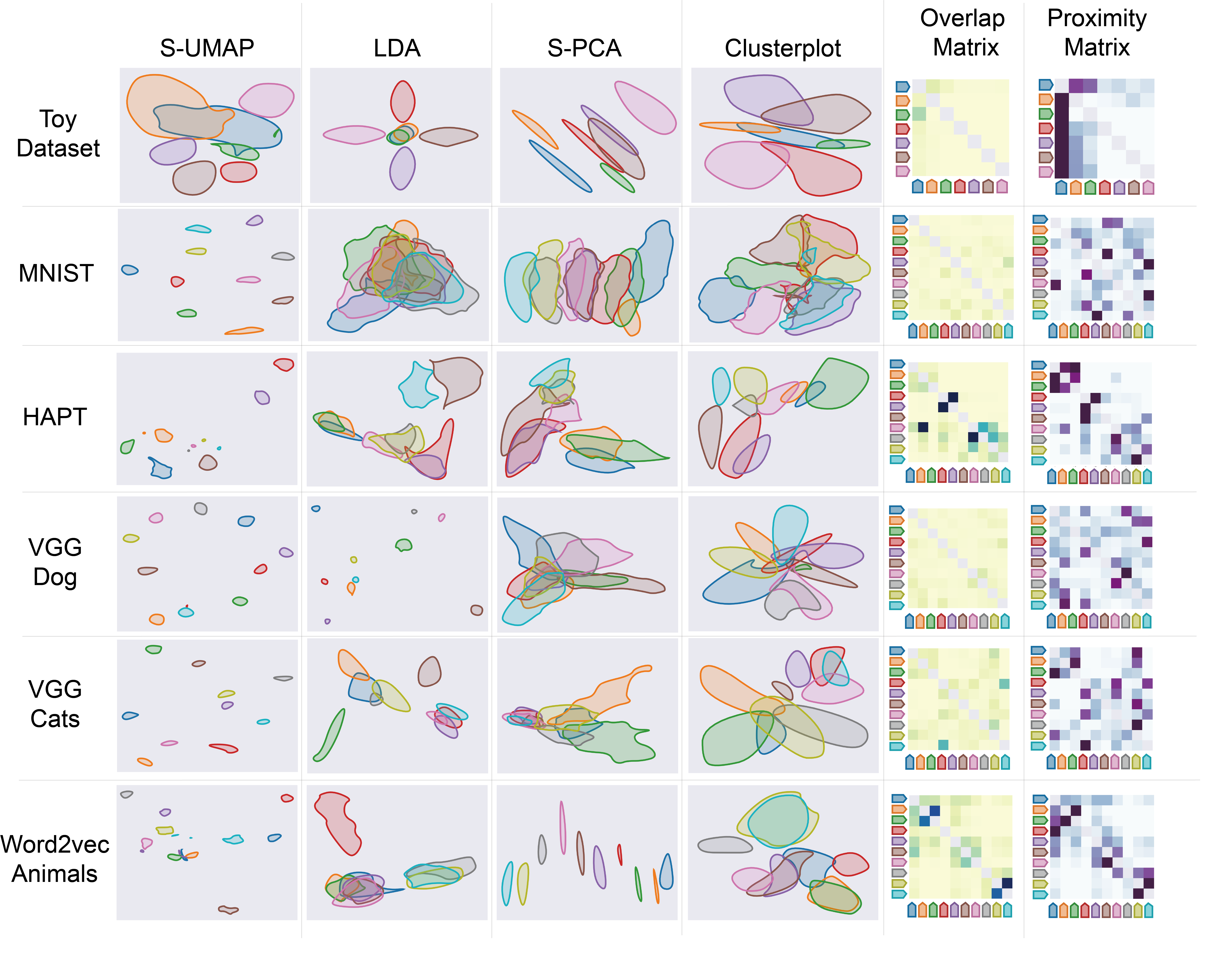}
    \caption{Different Supervised Algorithms over Several Datasets.}
  \label{img:compare_table}
\end{figure*}

 In this section, we demonstrate several examples of using our clusterplot to display multi-class high-dimensional data. Figure~\ref{img:compare_table} compares the different supervised algorithms of several datasets, including Supervised UMAP (S-UMAP), LDA, Supervised PCA (S-PCA) and Clusterplots. As can be seen, clusterplots outperform the other methods in providing insights of inter-class relationships. In the following, we elaborate on the performance of clusterplots in these examples and specifically compare  clusterplot to UMAP, which is an advanced version of t-SNE, and we use it as our dimensionality reduction backbone.
 
\begin{figure}[h!]
\centering
  \begin{subfigure}[t]{0.78\columnwidth}
    \includegraphics[width=\linewidth]{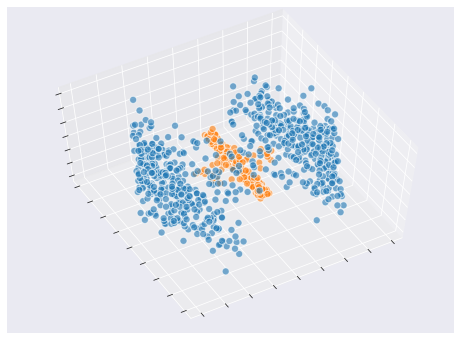}
    \caption{3D Dataset}
  \end{subfigure}
  
  
  \begin{subfigure}[t]{0.38\columnwidth}
    \includegraphics[width=\linewidth]{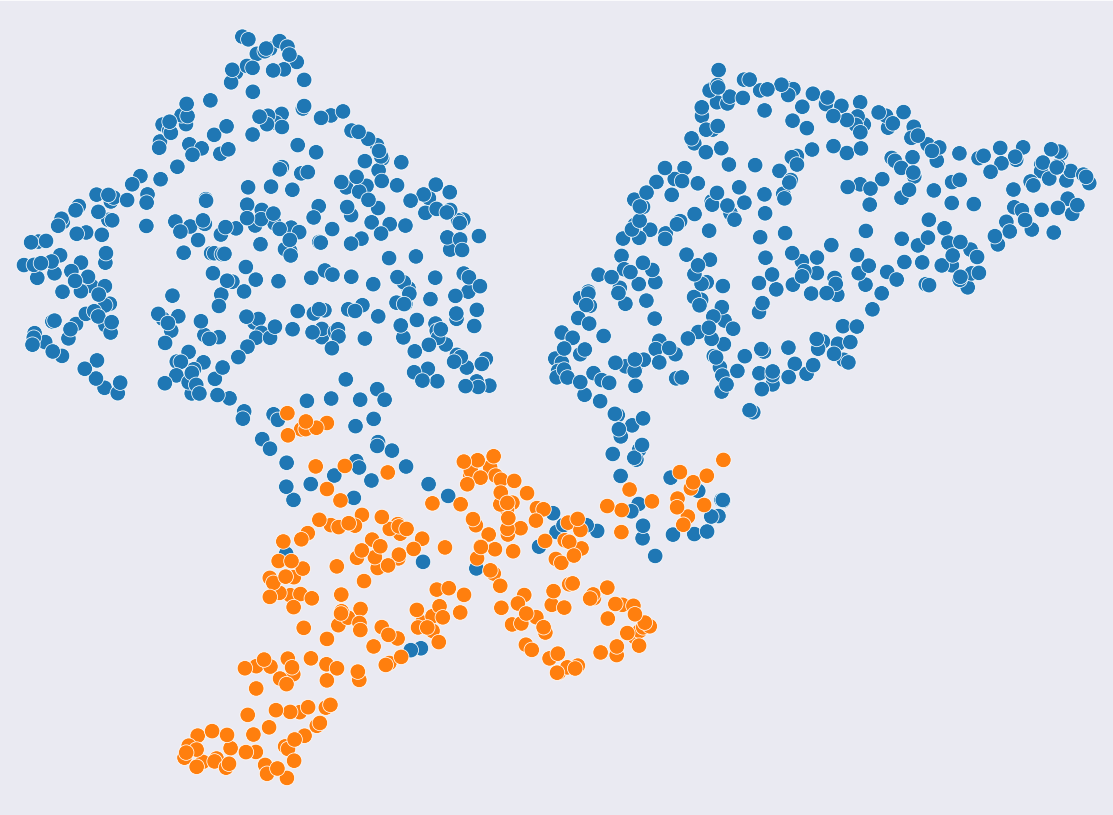}
    \caption{UMAP}
  \end{subfigure}
  \begin{subfigure}[t]{0.38\columnwidth}
    \includegraphics[width=\linewidth]{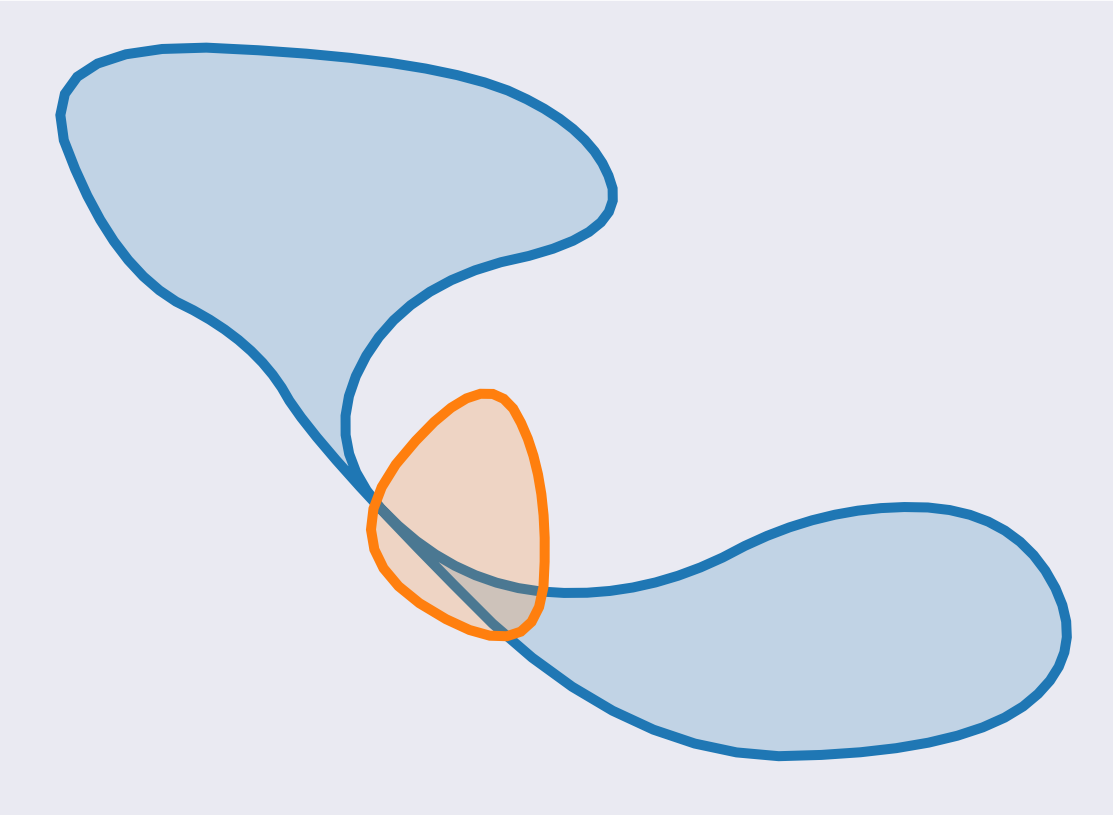}
    \caption{Clusterplot}
  \end{subfigure}
  
  \caption{UMAP and Clusterplot on simple 3D dataset with two classes with non-uniform density.}

  \label{fig:Hourglass}
\end{figure}

\begin{figure}[h!]
\begin{center}
    \begin{subfigure}[t]{0.75\columnwidth}
    \includegraphics[width=\linewidth]{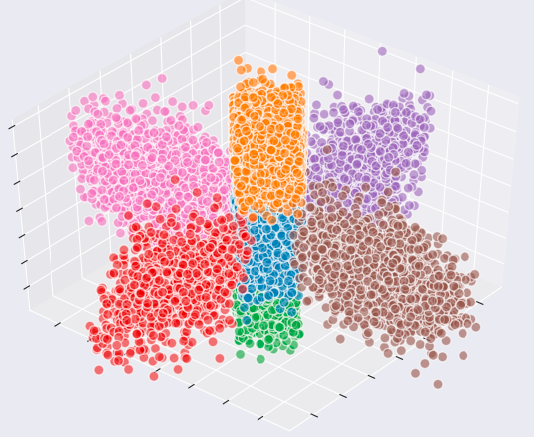}
    \caption{3D}
   \end{subfigure}
  
    \begin{subfigure}[t]{0.45\columnwidth}
    \includegraphics[width=\linewidth]{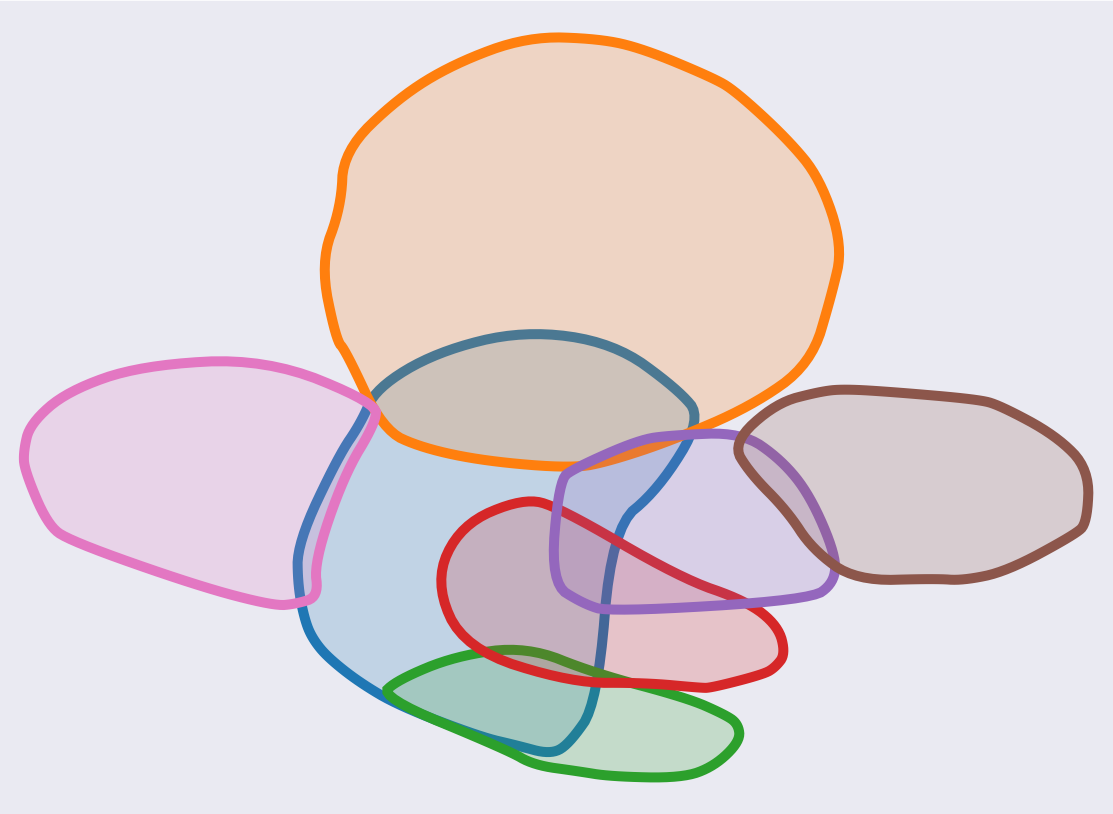}
    \caption{UMAP}
  \end{subfigure}
  \begin{subfigure}[t]{0.45\columnwidth}
    \includegraphics[width=\linewidth]{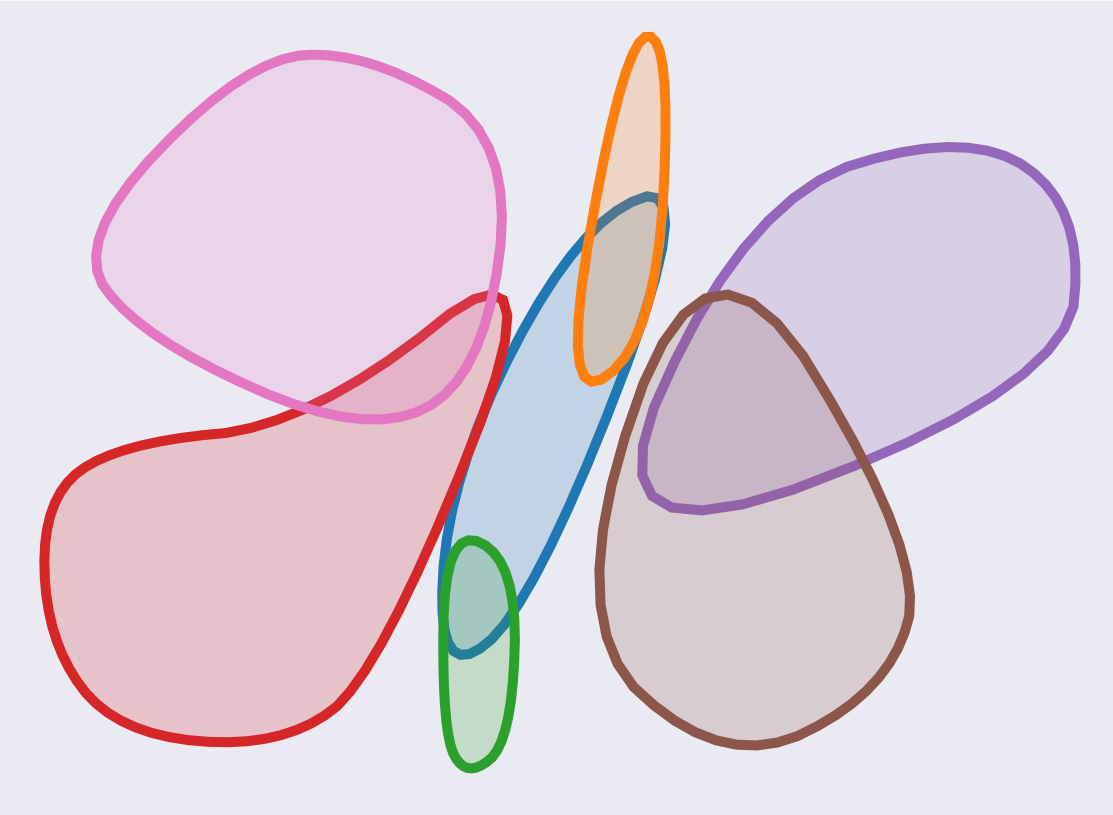}
  \caption{Clusterplot}
  \end{subfigure}
  
  \caption{UMAP and Clusterplot on 3D cross with non-uniform densities. The clusters of UMAP are displayed using our blobs.}
  \label{fig:Cross}
\end{center}
\end{figure}

\textbf{Toy Datasets} We start with two simple examples, both in 3D since in higher dimensions, we have no intuitive understanding of the inter-class relations. Furthermore, for clarity, the first example includes two classes only. See Figure \ref{fig:Hourglass} where the dataset has two clusters. One cluster (in Blue) has a hourglass-like shape, with two dense regions at both ends, and a sparse one in the bottleneck. The other cluster (in Orange), is denser and located close to the bottleneck. In (b) and (c), we compare side by side the 2D plots generated by UMAP and Clusterplot, respectively. We argue that the clusterplot significantly better represents these 3D clusters in 2D, both in terms of their relative sizes, and their global relative relations. Note that in 3D the orange cluster is significantly smaller than the two parts of the blue cluster. However in 2D, UMAP scales up the orange cluster relative to the blue one, while in the clusterplot, their relative sizes are preserved.
The second example, shown in Figure \ref{fig:Cross} consists of a 3D cross with seven clusters, each cluster with different density. Note that we display the clusters of UMAP using our blobs. This is to factoring out the effect of the means of visualizing the clusters. It can be seen that UMAP is struggling with the non-uniform clusters, and it spreads the blobs as the density increases. However, clusterplot handles the non-uniform densities, and the relative sizes of the clusters are better maintained in 2D.

We show the effectiveness of clusterplot in visually exploring high-dimensional data, quickly interpreting the inter-class relations, and expressing the confusion among the encoded data. In the following, we first trained a classifier on the data, then we calculated the confusion matrix of the classifier as means to measure inter-class relations. A confusion matrix indicates the likelihood of a classifier to succeed in its task on an unseen data point. Although confusion matrix is closely related to overlap matrix, we do not directly optimize it to generate the plots.

\begin{figure}[h!]
\centering
  \begin{subfigure}[t]{0.48\columnwidth}
    \includegraphics[width=\linewidth]{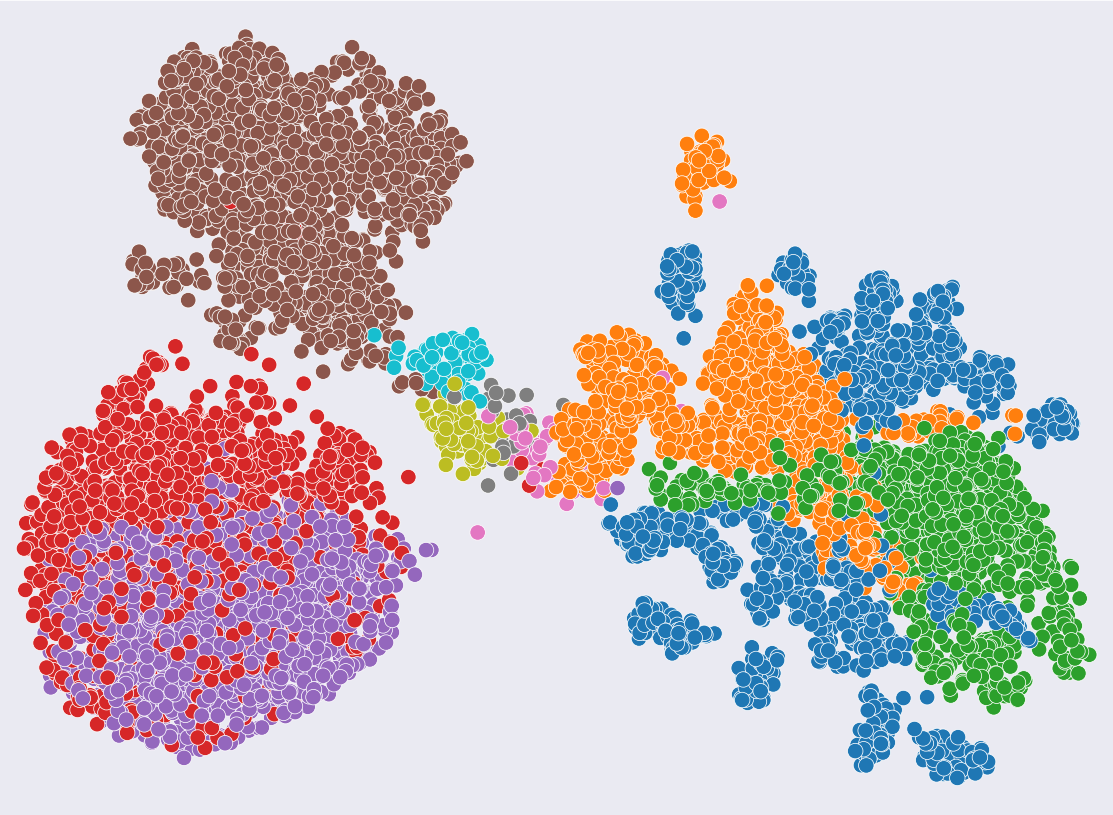}
    \caption{UMAP}
  \end{subfigure}
  \begin{subfigure}[t]{0.48\columnwidth}
    \includegraphics[width=\linewidth]{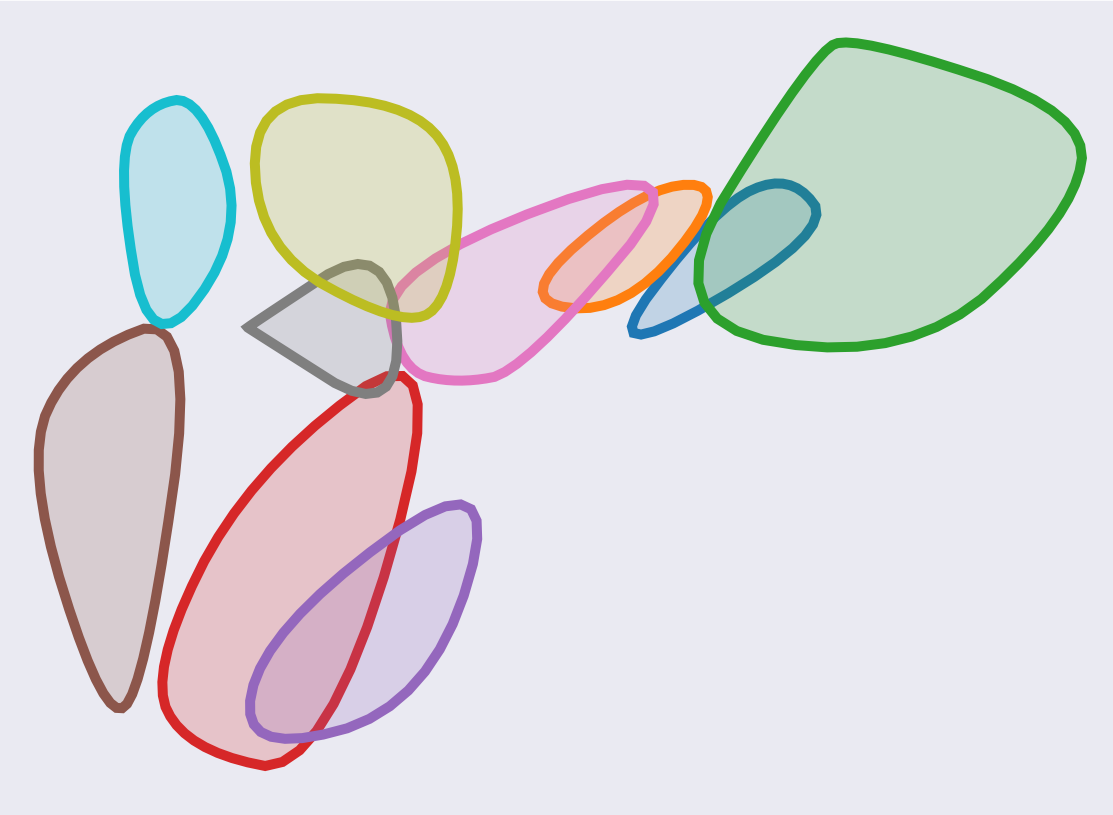}
    \caption{Clusterplot}
  \end{subfigure}
  
  \begin{subfigure}[t]{0.48\columnwidth}
    \includegraphics[width=\linewidth]{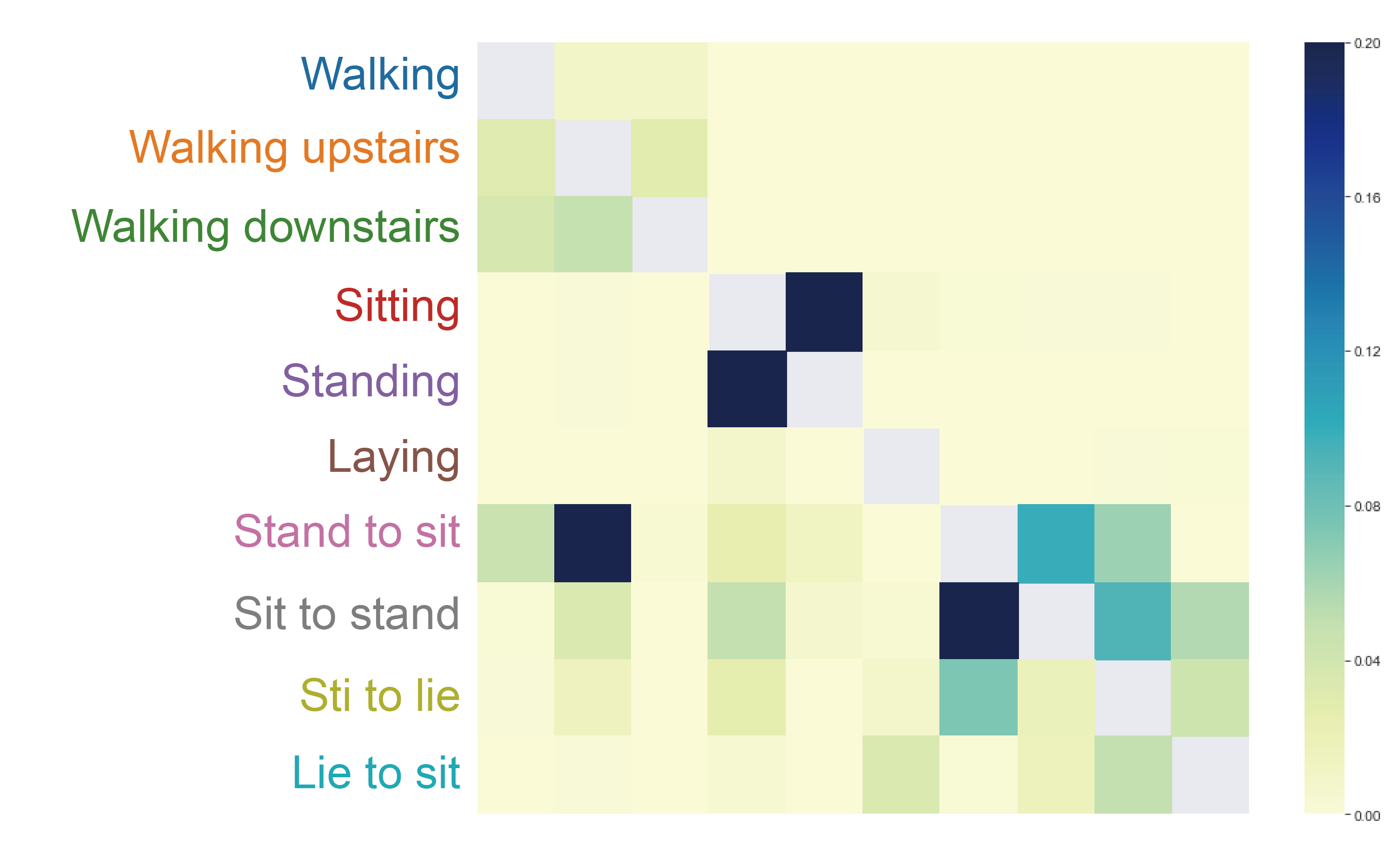}
    \caption{High-Dim Overlap}
  \end{subfigure}
  \begin{subfigure}[t]{0.48\columnwidth}
    \includegraphics[width=\linewidth]{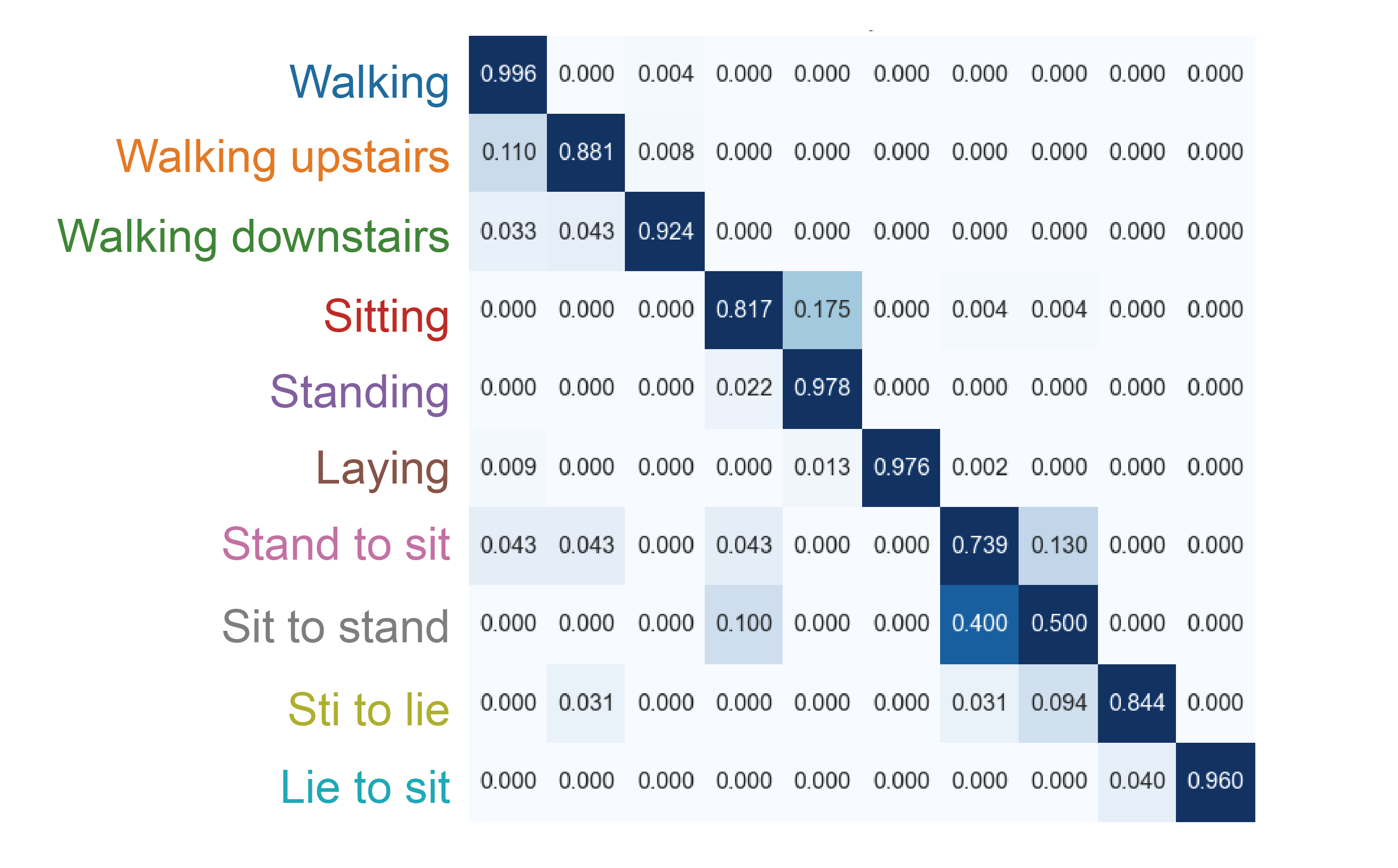}
    \caption{Confusion Matrix}
  \end{subfigure}
  
  \caption{HAMP: UMAP, Clusterplot, Overlap and confusion matrices of HAPT dataset. Blue: Walking, orange: walking upstairs, green: walking downstairs, red: sitting, purple: standing, brown: laying, pink: stand to sit, grey: sit to stand, pear: sit to lie, light blue: lie to sit.}

  \label{fig:HAPT}
\end{figure}

\textbf{HAPT Dataset} Our third example is a real dataset of Human Activities Postural Transitions (HAPT)~\cite{reyes2016transition}. This dataset contains features extracted from smartphones, while the people performed activities such as sitting and walking. Figure \ref{fig:HAPT} shows UMAP, clusterplot and confusion matrix of a network based classifier trained on this dataset. Note that the measured overlap is correlated with the confusion of the classifier as can be observed, for example, between the class \textit{sit-to-stand} to class \textit{stand-to-sit}. 
In addition, the overlap between the classes are better perceived in the clusterplot than in the UMAP one. For example, \textit{walking} (blue) is overlapped with both \textit{walking-upstairs} (orange) and \textit{walking-downstairs} (green), however there is no overlap between \textit{walking} and \textit{sitting} (red).

\begin{figure}
\begin{center}
  \begin{subfigure}[t]{0.90\columnwidth}
    \includegraphics[width=\linewidth]{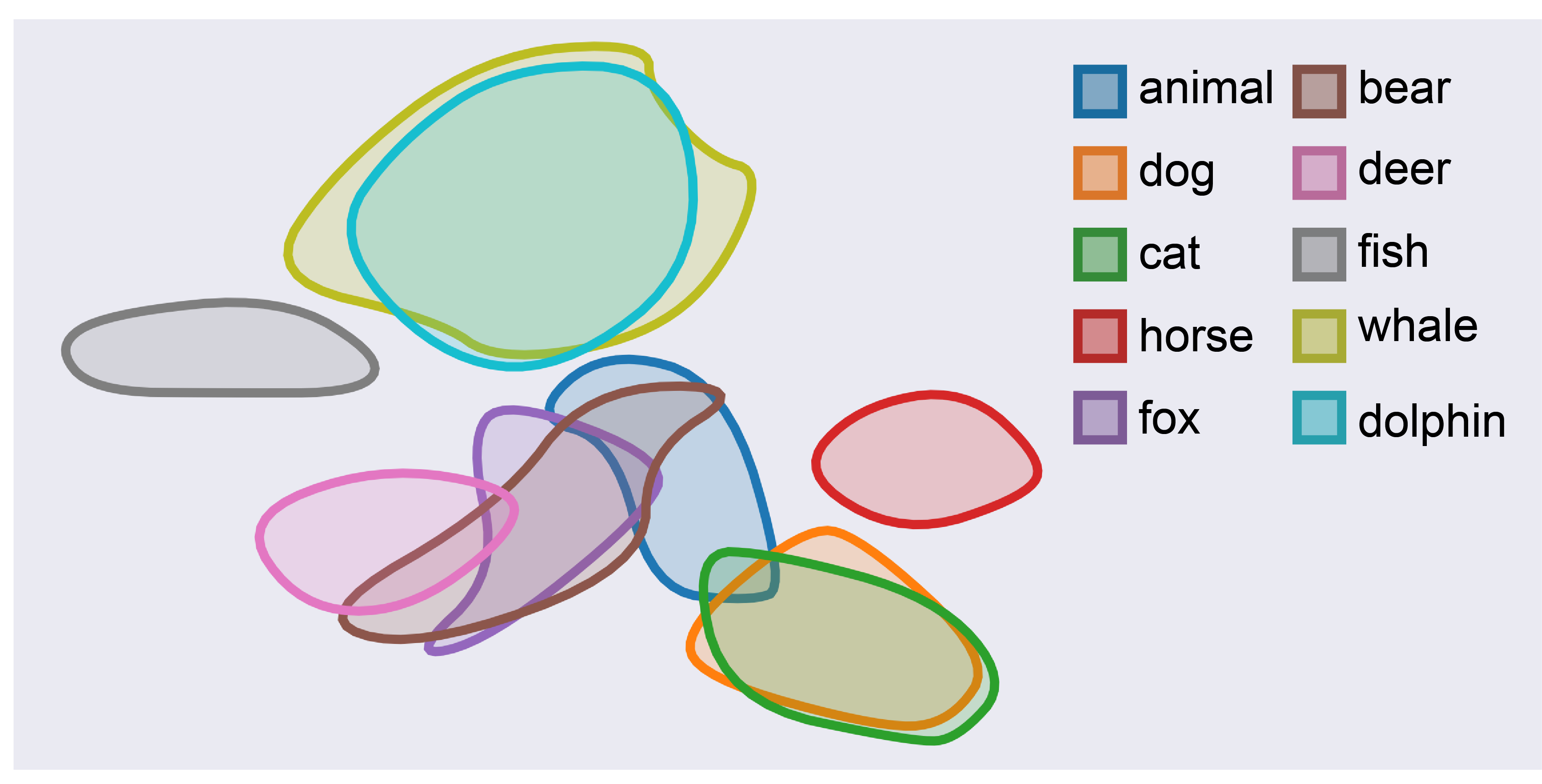}
    \caption{Clusterplot}
  \end{subfigure}
  \begin{subfigure}[t]{0.45\columnwidth}
    \includegraphics[width=\linewidth]{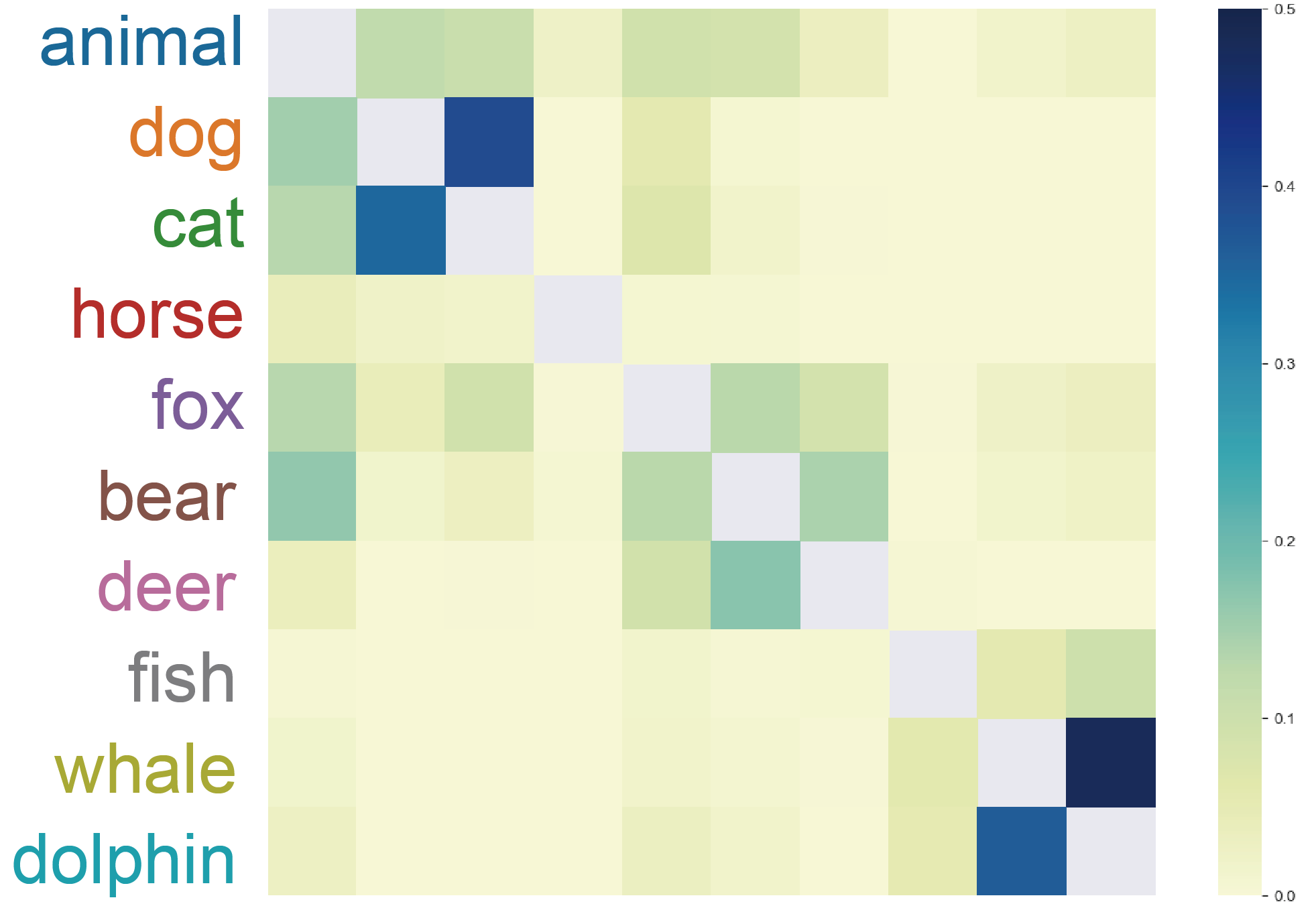}
    \caption{Overlap}
  \end{subfigure}
  \begin{subfigure}[t]{0.45\columnwidth}
    \includegraphics[width=\linewidth]{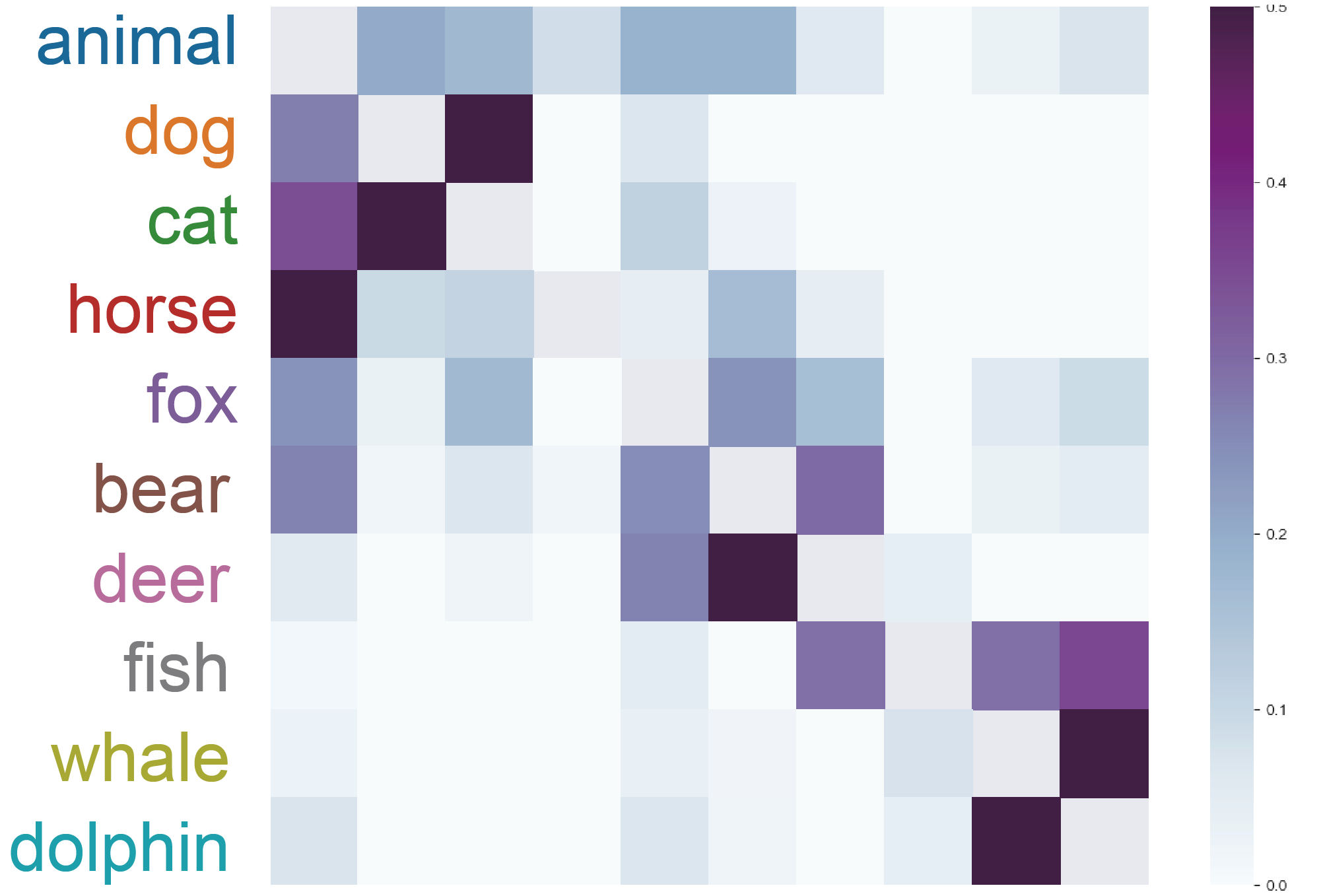}
    \caption{Proximity}
  \end{subfigure}
  \caption{Clusterplot, overlap and proximity matrices measured on words embedding of word related to animals}
  \label{fig:word2vecAnimals}
\end{center}
\end{figure}

\textbf{Word2Vec Animal Dataset} This example shows clusterplots with a more complex dataset. We collected vector representation of words obtain by word2vec Skip-gram  model\cite{mikolov2013distributed}, and used the 300-dimensional pre-trained word embedding by Google, trained on Google News dataset. We chose ten words that we used as labels, including animal, dog, cat, horse, fox, bear, deer, fish, whale and dolphin. Then, for each label we choose the 100 most similar words representations to the label representation based on cosine-similarity, each chosen word is a labeled sample. By the definition of Skip-gram, words with similar context are close to each other in the embedding space, it can be easily seen in Figure~\ref{fig:word2vecAnimals} that \textit{dog} and \textit{cat} are overlapped and also \textit{whale} and \textit{dolphin} are overlapped. In addition, the sea-animals, \textit{fish}, \textit{whale} and \textit{dolphin} are close to each other, and the land-wild-animals, \textit{dear}, \textit{fox} and \textit{bear} are also close to each other. Note that \textit{animal} blob is located in the center and the blobs of the types of animals are surrounding it.

\section{Usage Scenarios in Deep Features}

\begin{figure}[h!]
\begin{center}
  \begin{subfigure}[t]{0.32\columnwidth}
    \includegraphics[width=\linewidth]{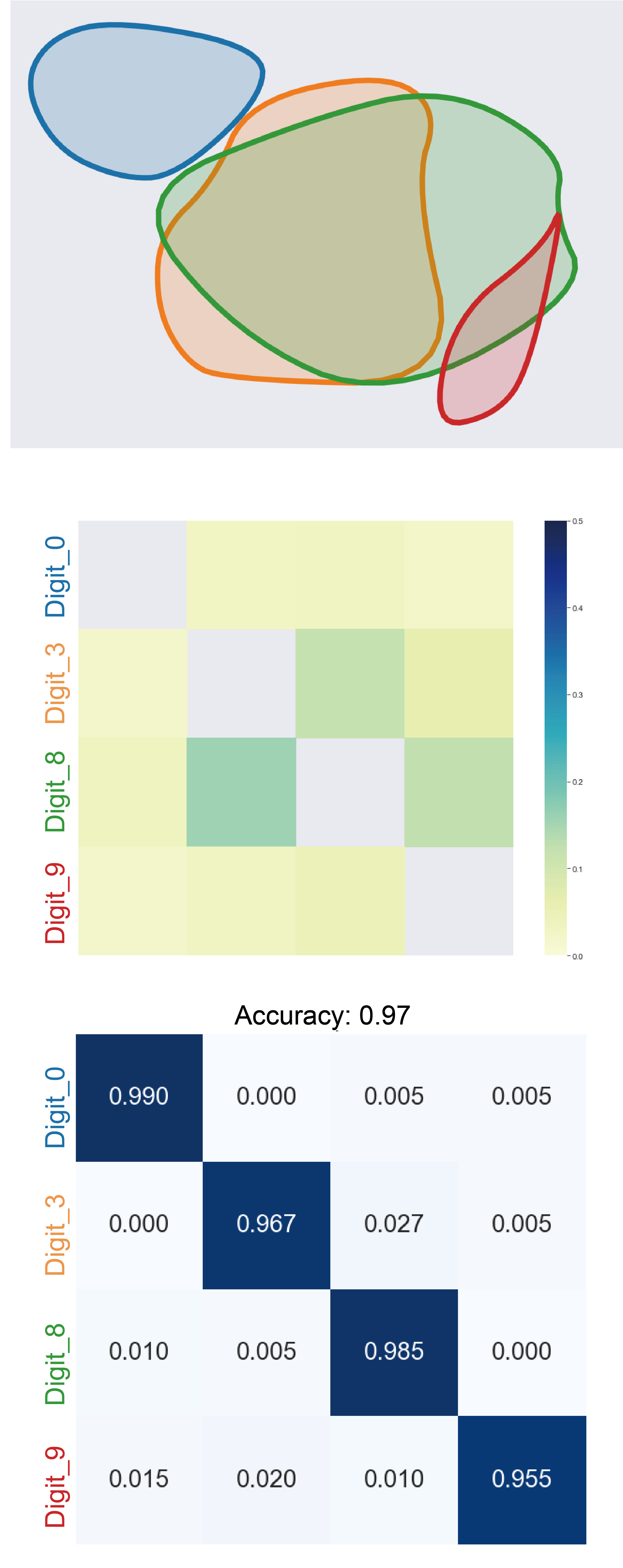}
    \caption{Orig-Dim 784}
  \end{subfigure}
  \begin{subfigure}[t]{0.32\columnwidth}
    \includegraphics[width=\linewidth]{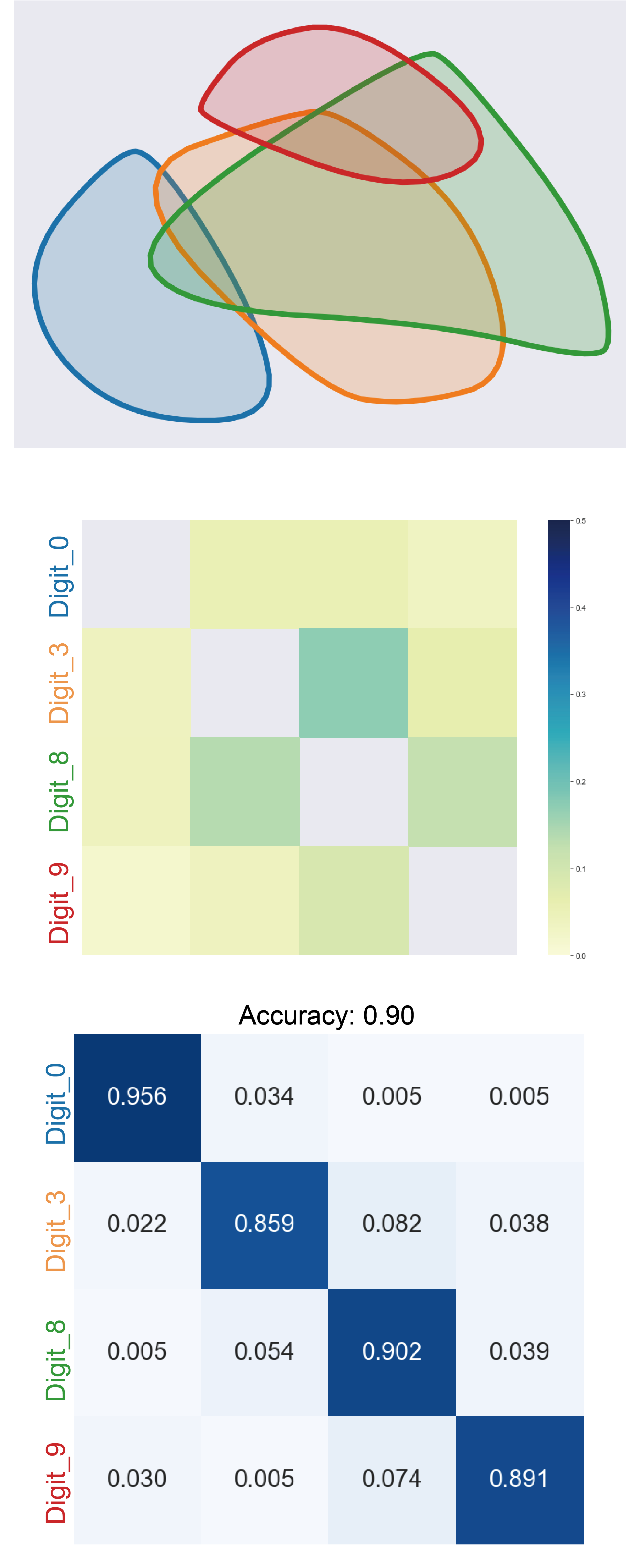}
    \caption{AE Dim 32}
  \end{subfigure}
  \begin{subfigure}[t]{0.32\columnwidth}
    \includegraphics[width=\linewidth]{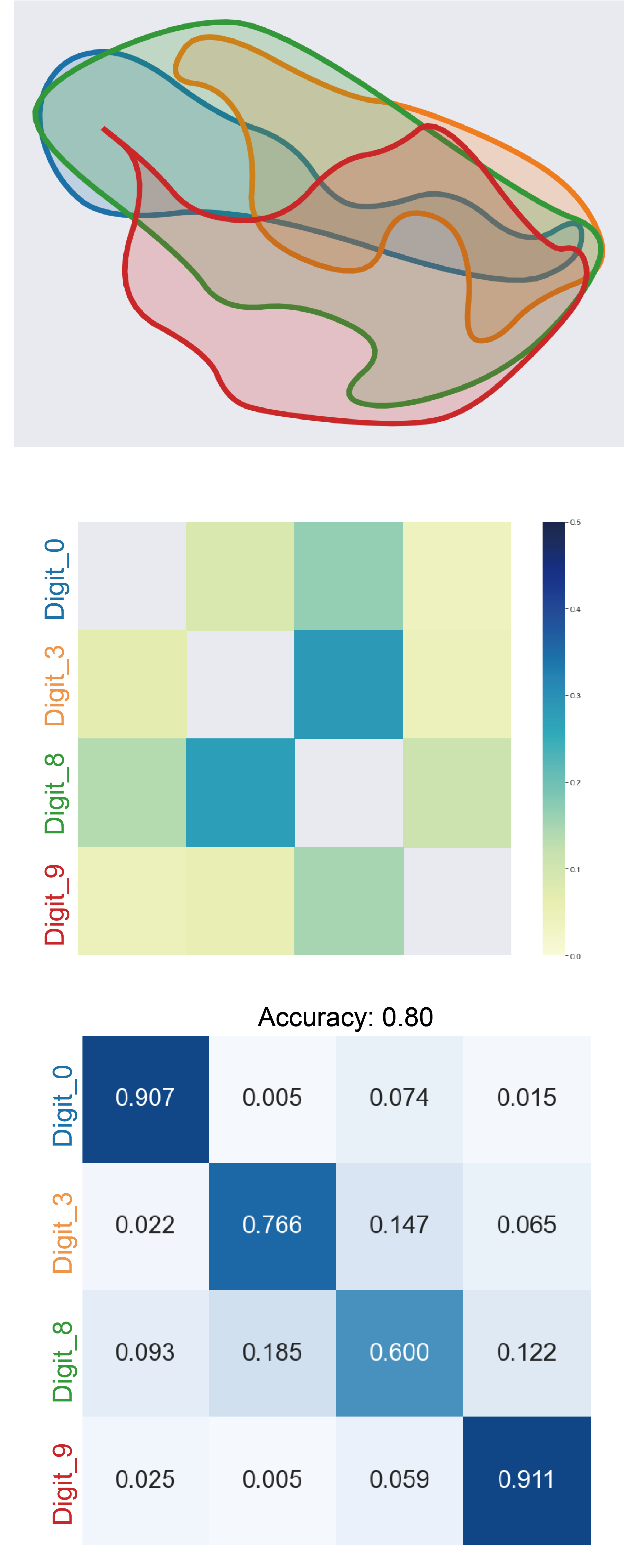}
  \caption{AE Dim 4}
  \end{subfigure}
  
  
  
  \caption{Clusterplot, High-Dim overlap matrix and confusion matrix of four digits of MNIST (Blue: 0, Orange: 3, Green: 8, Red: 9), 32-dim and 4-dim features as extracted from hidden layer of AE.}
  \label{fig:MNIST4_AE}
\end{center}
\end{figure}

In the following, we demonstrate a number of usage scenarios of our method as means to explore the inter-relations among deep features generated by neural networks. 

\textbf{MNIST Dataset} Let us start off with a trained Auto Encoder (AE) of an MNIST dataset, which is a simple set of images of the letters $0$ to $9$. We trained an AE with two different bottleneck sizes: 32-dim and 4-dim. 
Figure \ref{fig:MNIST4_AE} shows the clusterplot, and the overlap and confusion matrices of the deep features generated by the AEs. 
Clearly, the deep features generated with the overly narrow bottleneck of $4$ yields much more confusion by the classifier. Furthermore, Figure \ref{fig:MNIST4_AE} shows the correlation between the overlap matrices and the confusion matrices. For example, it can be easily seen that the overlap between the digit eight to three and zero in dimension 4 is high, and respectively, the confusion matrix shows that the classifier performs poorly on theses digits.



\begin{figure*}[h!]
\begin{center}
  \begin{subfigure}[t]{0.23\linewidth}
    \includegraphics[width=\linewidth]{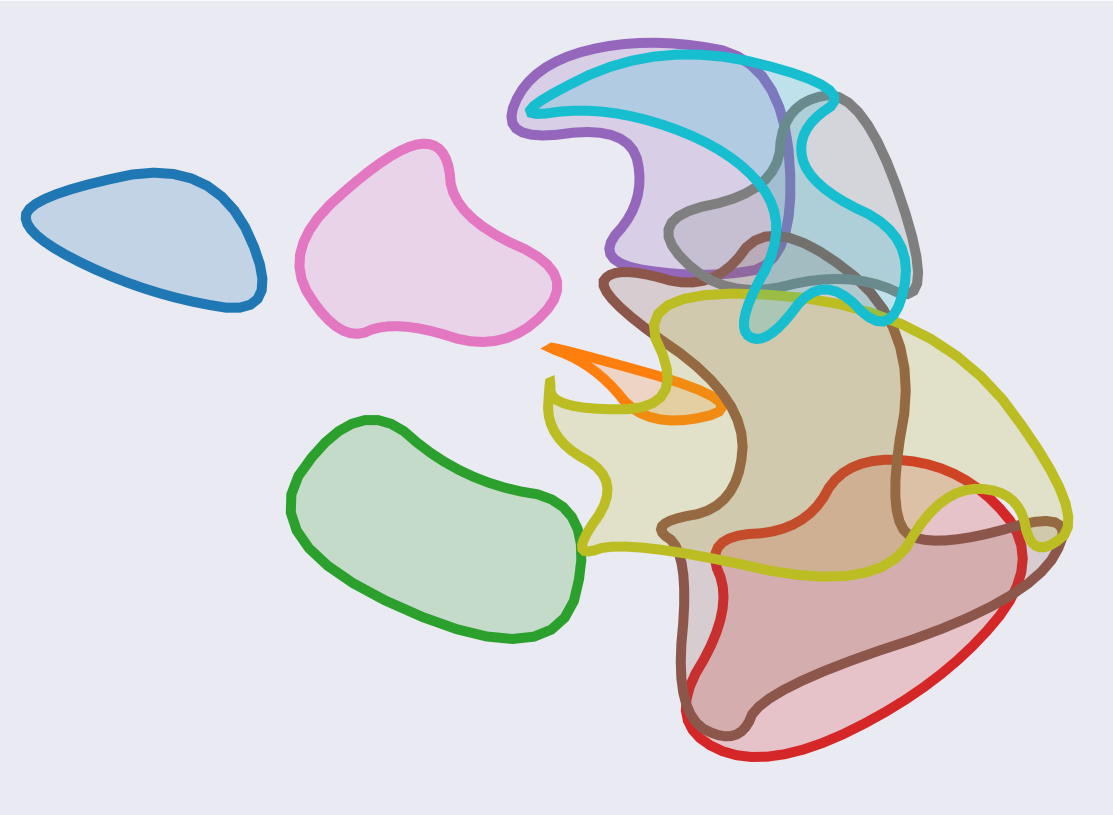}
    \caption{80\% accuracy}
  \end{subfigure}
  \begin{subfigure}[t]{0.23\linewidth}
    \includegraphics[width=\linewidth]{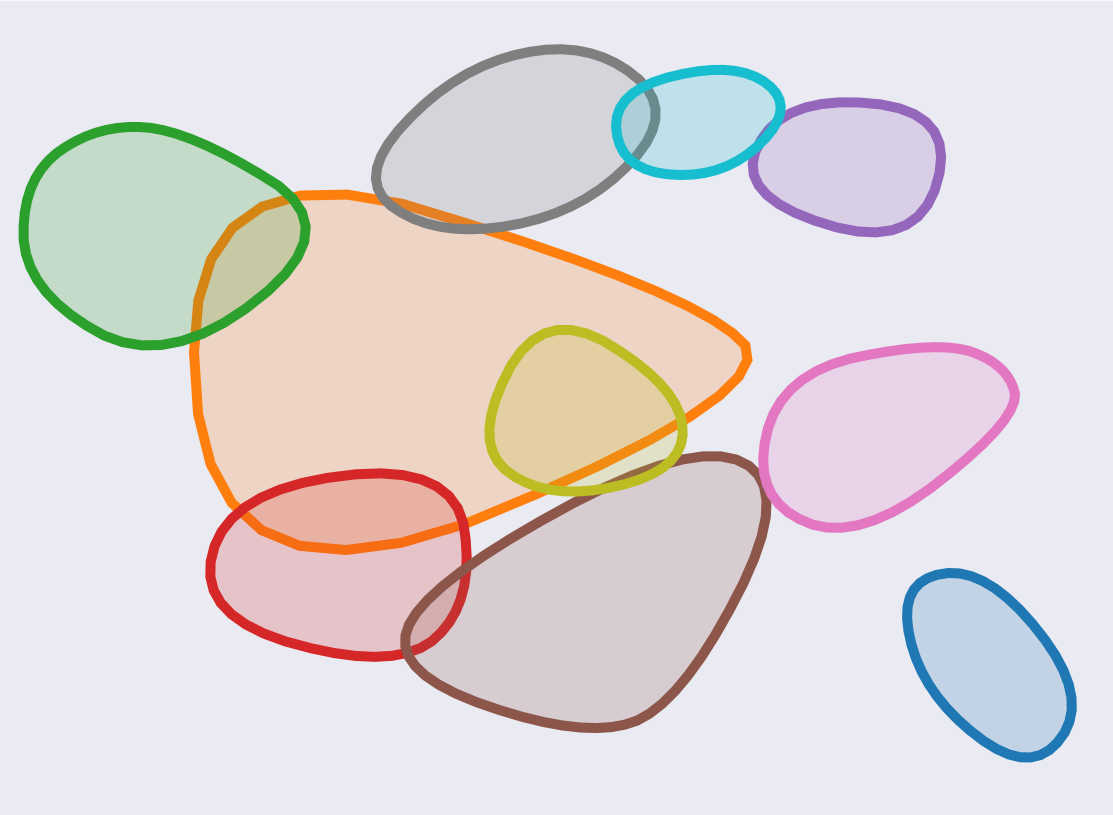}
    \caption{94\% accuracy}
  \end{subfigure}
  \begin{subfigure}[t]{0.23\linewidth}
    \includegraphics[width=\linewidth]{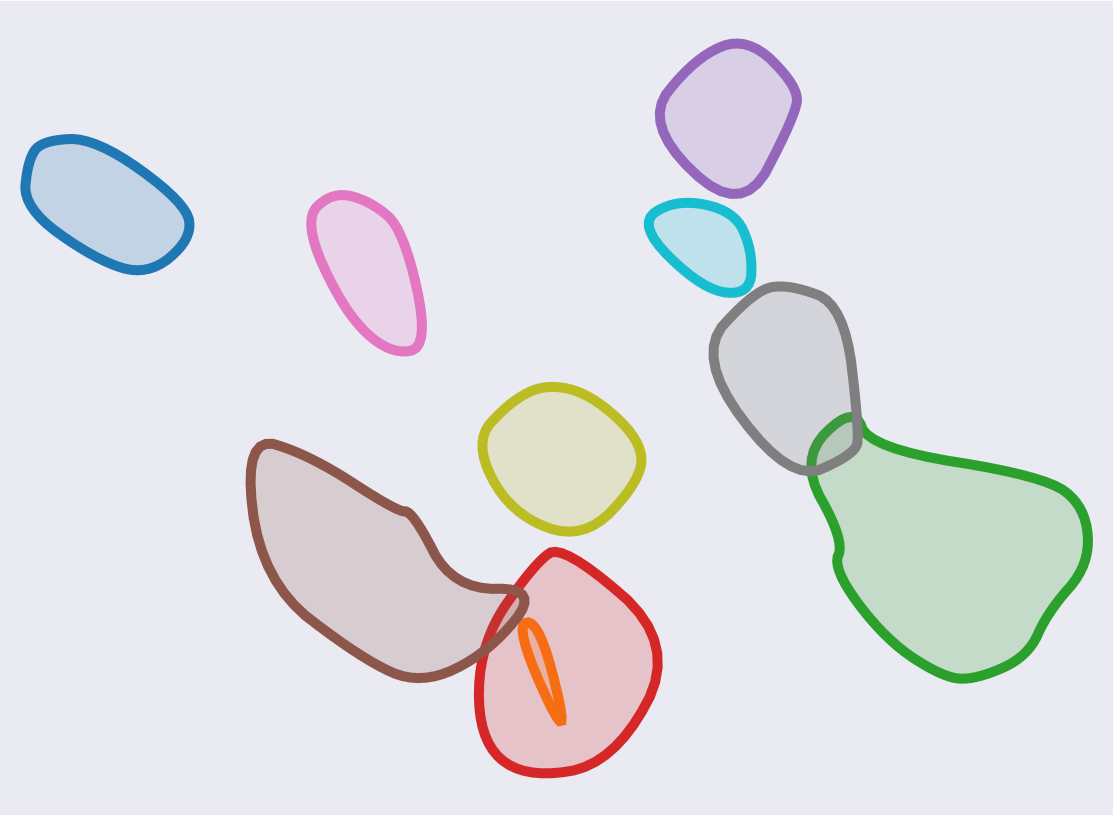}
    \caption{96\% accuracy}
  \end{subfigure}
    \begin{subfigure}[t]{0.23\linewidth}
    \includegraphics[width=\linewidth]{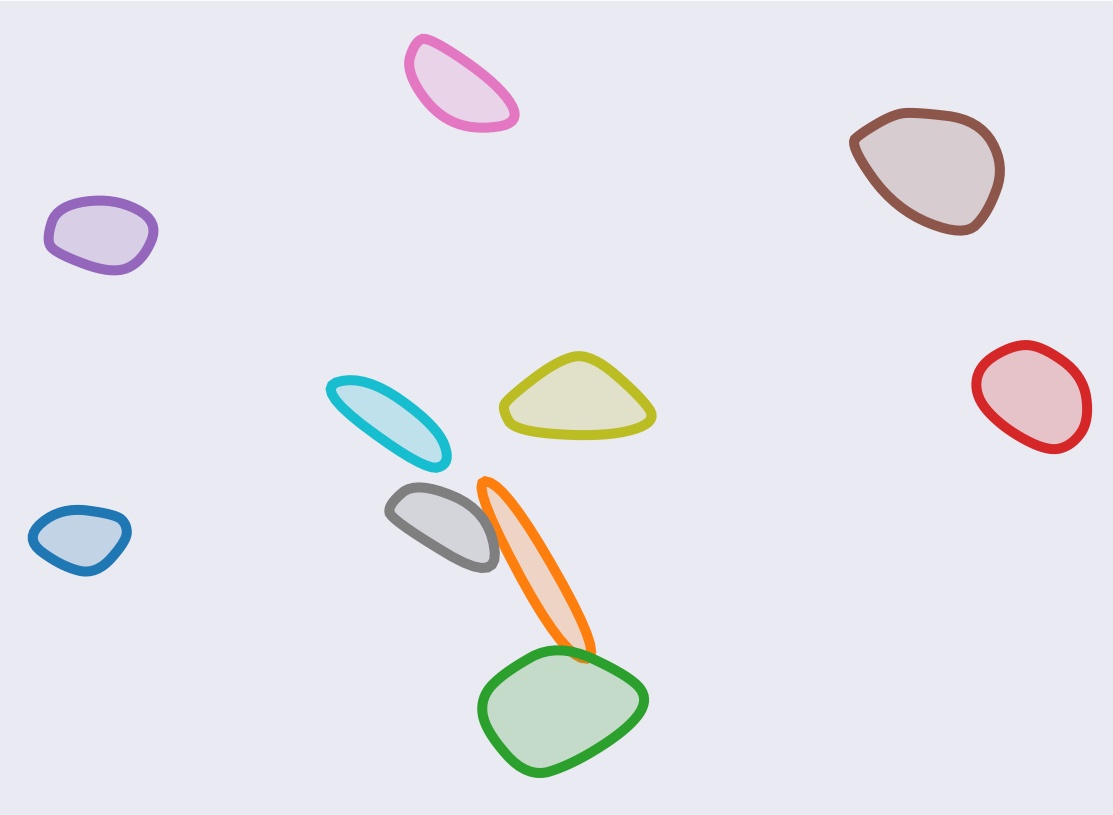}
    \caption{98.5\% accuracy}
  \end{subfigure}
  
  \caption{Feature vector (dim-128) extracted by deep convolution neural network at different stages of the training on MNIST dataset. }
  \label{fig:MNISTDeepFull}
\end{center}
\end{figure*}

To further demonstrate the usefulness of clusterplot, we trained deep convolution network on the MNIST dataset. We evaluated the accuracy of the network at the different stages of the training, and extracted the outputs from the last fully connected layer on the test set. Figure \ref{fig:MNISTDeepFull} shows clusterplot of the extracted features at different stages of the training. It can be seen that as the training progresses, the overlapping among the clusters is reduced as expected, where finally, in the last plot, where the accuracy is 98.5\%, the clusters are almost completely separated.

\begin{figure*}[!htb]
\begin{center}
  \begin{subfigure}[t]{0.95\linewidth}
    \includegraphics[width=\linewidth]{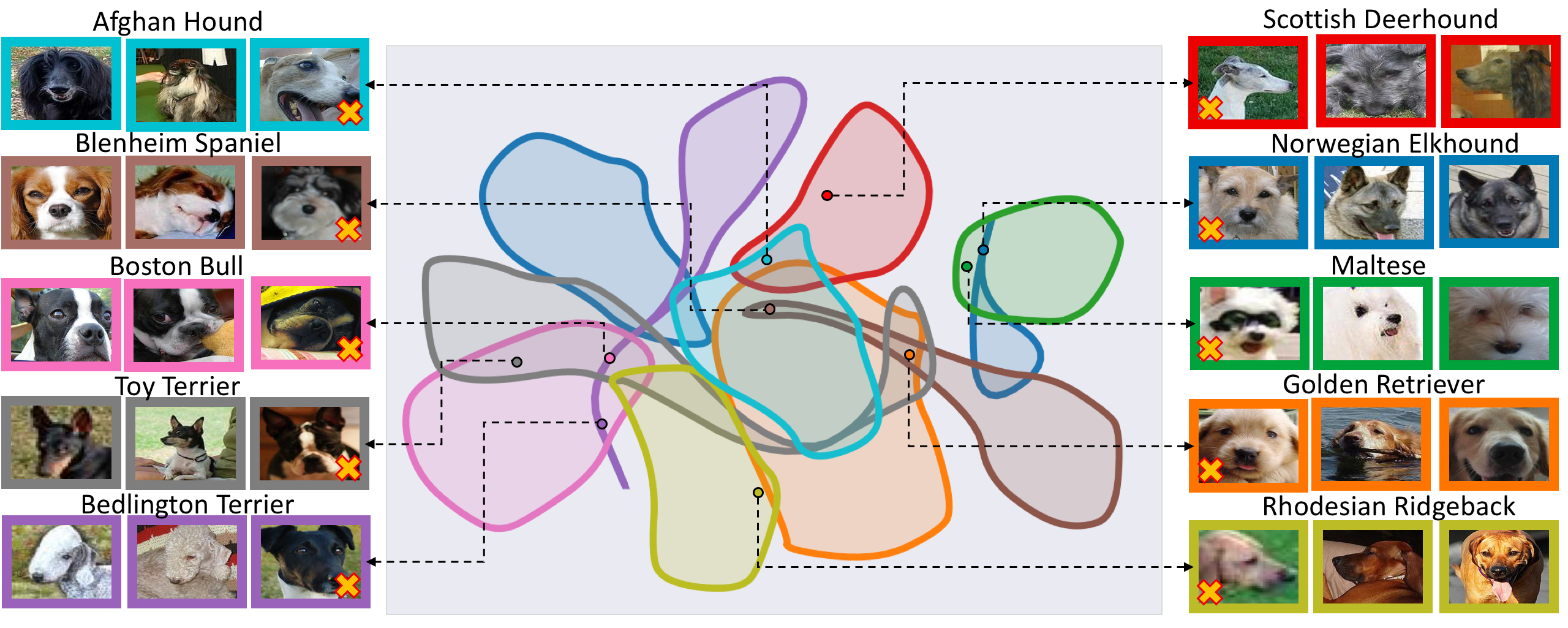}
    \caption{Features of MobileNet deep neural network on images of 10 classes of dogs from Stanford Dogs Dataset}
  \end{subfigure}
  \begin{subfigure}[t]{0.95\linewidth}
    \includegraphics[width=\linewidth]{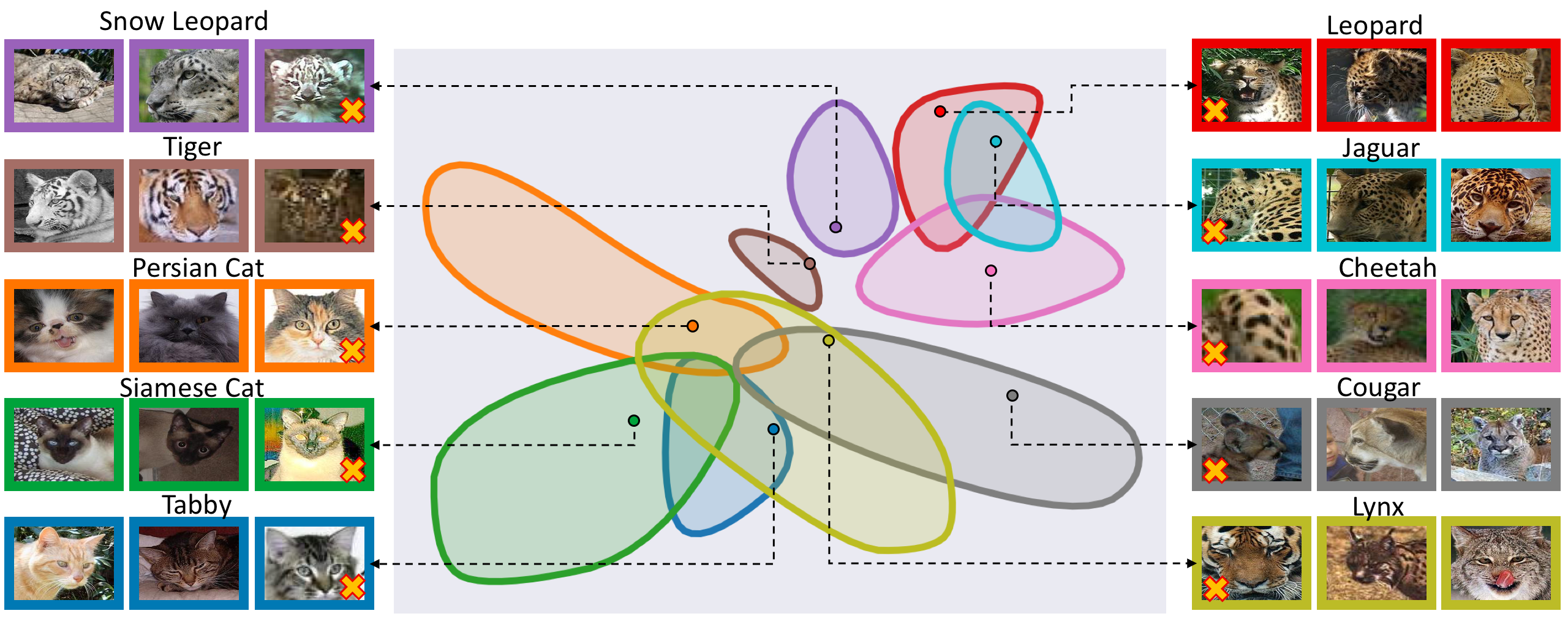}
    \caption{Features of VGG deep neural network on images of 10 classes of cats from ImageNet Dataset}
  \end{subfigure}
  \caption{Features of popular deep neural network on images from Stanford Dogs  and ImageNet datasets. Images on the margins were chosen using KNN-classifier. The images with yellow $X$ symbols are misclassified, and the origin of the arrows is their location in the embedded space.}
  \label{fig:MobileNetDogsAnnot}
\end{center}
\end{figure*}

\textbf{VGG Dogs Dataset} Figure \ref{fig:MobileNetDogsAnnot}(a) shows clusterplot of features of MobileNet deep neural network on images from 10 classes of the Stanford Dogs Dataset.
The network extracts features of dimension 1281 from the images. The images on the margins were chosen using a KNN classifier; images with yellow $X$ symbols are misclassified, while images without it are classified to their correct classes. The origin of the arrows is the embedded position of the misclassified images. 

Note that the overlap between the classes is visually expressed in the clusterplot, while the locations of the misclassified images emphasizes the potential confusion between the classes. For instance, the misclassified \textit{Norwegian Elkhound} is overlapped with the \textit{Maltese} dog and it is clear that this dog is more similar to \textit{Maltese} rather than \textit{Norwegian Elkound}. The misclassified \textit{Golden Retriever} is a puppy that is also overlapped with the \textit{Blenheim Spaniel} which is a small dog. Other misclassifed images seem like images with incorrect labels in the dataset, for example the \textit{Bedlington Terrier} looks like a \textit{Toy Terrier}, and the \textit{Toy Terrier} looks like \textit{Boston Bull}.

\begin{figure*}[!htb]
\begin{center}
  \begin{subfigure}[t]{0.32\linewidth}
    \includegraphics[width=\linewidth]{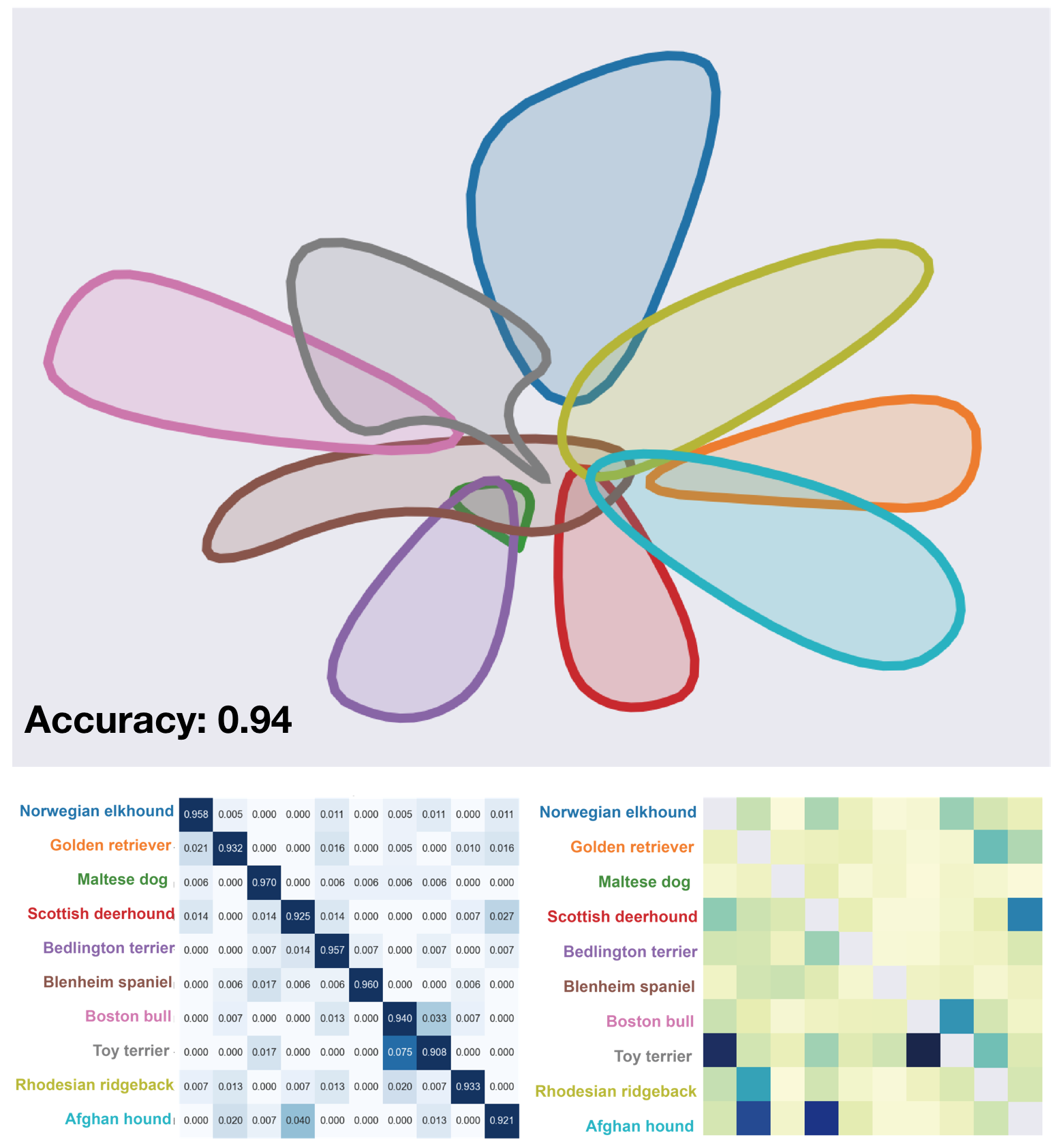}
    \caption{VGG-Dim 4096}
  \end{subfigure}
  \begin{subfigure}[t]{0.32\linewidth}
    \includegraphics[width=\linewidth]{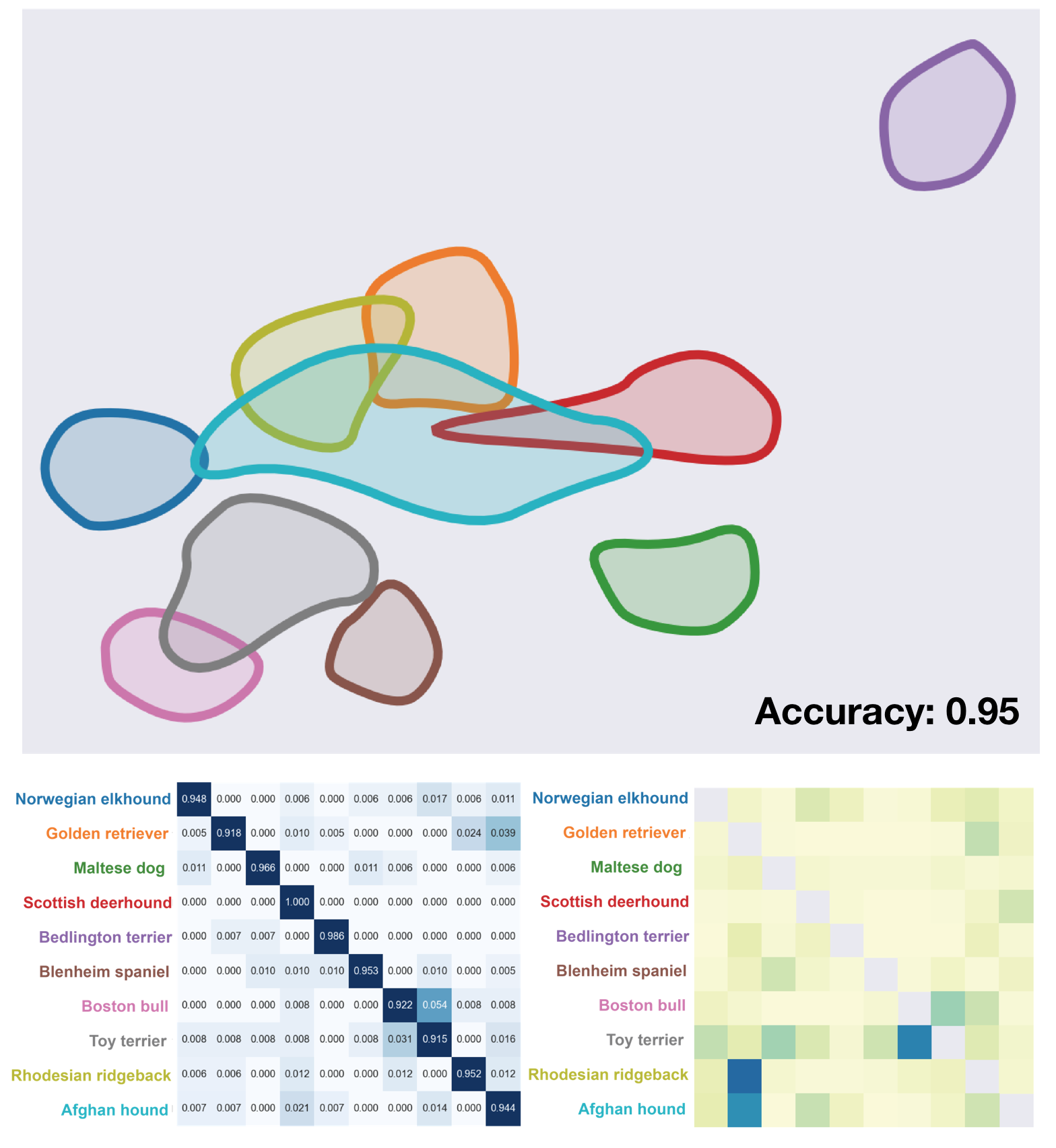}
    \caption{DenseNet-Dim 1025}
  \end{subfigure}
  \begin{subfigure}[t]{0.32\linewidth}
    \includegraphics[width=\linewidth]{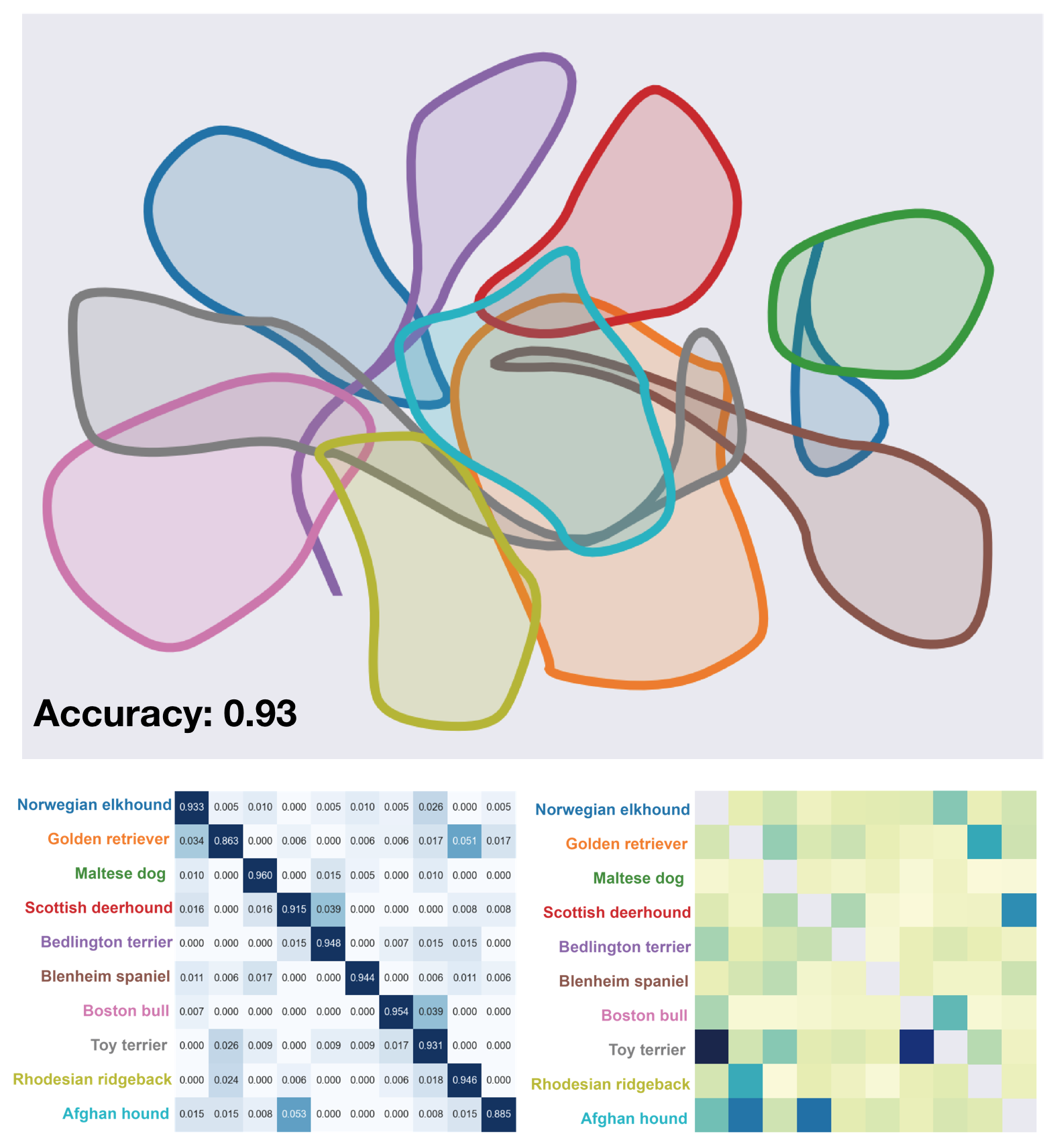}
  \caption{MobileNet-Dim 1281}
  \end{subfigure}
  
  
  
  \caption{Clusterplot, High-Dim overlap matrices and confusion matrices of features generated by applying three common deep neural networks on the same classified images from Stanford Dogs Dataset. Blue: Norwegian
Elkhound, orange: Golden Retriever, green: Maltese, red: Scottish Deerhound, purple: Bedlington Terrier, brows: Blenheim Spaniel, pink: Boston Bull, grey: Toy Terrier, pear:
Rhodesian Ridgeback, light blue: Afghan Hound}
  \label{fig:DeepFeaturesComparison}
\end{center}
\end{figure*}

In addition to the MobileNet features, we evaluated the extracted features of Stanford Dogs Dataset from VGG and DenseNet networks, which have dimensions 4096 and 1025 respectively. Figure \ref{fig:DeepFeaturesComparison} shows the corresponding clusterplot, High-dim overlap and confusion matrices of corresponding features of these networks. It seems like the features of the three networks have an overlap between \textit{Rhodesian Ridgeback} and \textit{Afghan Hound} to \textit{Golden Retriever} and the corresponding confusion matrices have entries that reconfirm that.
The overlap between \textit{Rhodesian Ridgeback} and \textit{Afghan Hound} and the \textit{Golden Retriever} can be easily observed in the VGG clusterplot, as the \textit{Golden Retriever} lays just between the \textit{Rhodesian Ridgeback} and the \textit{Afghan Hound} regions in the plot. 
However in DenseNet and MobileNet clusterplot, the \textit{Afghan Hound} blob is overlapped with all the other classes. It seems that in these cases, clusterplot struggles with the \textit{Afghan Hound} cluster.
Furthermore, note that the correlation between the overlap and the confusion matrix can also be seen between the \textit{Boston Hull} and the \textit{Toy Terrier}. 

\textbf{VGG Cats Dataset} We applied similar evaluation on features (4096 dimensions) extracted by VGG on 10 classes of cats from ImageNet dataset \cite{imagenet_cvpr09}. Figure \ref{fig:MobileNetDogsAnnot}(b) demonstrates the clusterplot of these features. Also in this case, the overlap between the classes is clearly presented in the plot. For example between big cats like \textit{Cheetah}, \textit{Jaguar}, \textit{Leopard} and \textit{Snow-Leopard}, and also between the domestic cats like \textit{Persian-Cat}, \textit{Siamese-Cat} and \textit{Tabby}. Furthermore, the locations of the misclassified images help to understand the overlap and confusion between the classes. For example, the misclassified \textit{Persian-Cat} has similar colors to \textit{Siamese-Cat}, and it is located in the overlap between the classes. Also, the misclassified \textit{Lynx} looks exactly like a \textit{Tiger} and it is close to the \textit{Tiger} blob, this also may be an example of incorrect label in the dataset. Note also that the misclassified \textit{Cheetah} has no face at all.

\section{Evaluation}

To evaluate the performance of Clusterplots and their contribution to the perception the inter-relations among clusters, we conducted a formal user study, in which participants performed cluster-related tasks to compare the effectiveness and intuitiveness of Clusterplots versus a baseline matrix visualization tool~\cite{chen2002generalized}. A matrix form is considered as a conventional choice to visualize pair-wise information (see Figure~\ref{fig:teaser}). 
In our evaluation, we focus on the perception of overlapping among clusters. Thus, other supervised alternatives, like supervised-UMAP are irrelevant, since they do not aim to display pair-wise information.
The goal of the study was to examine whether Clusterplots enhance the perception of inter-cluster overlaps beyond that of a matrix visualization (hereafter refer it to Matrix for short). To better understand the effect of Clusterplots, given different cluster complexity levels, we examined another one independent variables that might affect the way users perceive cluster information, i.e., the number of clusters. 


\subsection{Methodology}

The experimental design was a $2 \times 3$ within-subject design
with two main independent variables: \textit{visualization type} (Clusterplots or Matrix) and
\textit{number of clusters} (4, 6 and 8). 
%
We examined three different user tasks, designed according to the set tasks defined in Set Visualization ~\cite{Alsallakh2014} adapted specifically to the analysis of intersection/overlap between sets. Due to the complexity of the experimental design, we analyze and report on the result of each task separately, and thus, do not include the tasks as an independent variable in our analysis. 

\paragraph{Participants} 20 participants were recruited from a local university (six females) with an average age of 23 (SD = 2). All were science engineering students. Nine of them were undergraduates and 11 were post-graduate students. 19 participants reported no prior or little visualization knowledge, and only one participants reported having a medium to high level of knowledge in visualization. All participants gave informed consent and the study conformed to the ethics procedure of our university.

\paragraph{Tasks} Three different tasks were used to evaluate the performance of Clusterplots:

\begin{itemize}
    \item Task 1: Does cluster \textit{A} overlap with cluster \textit{B}?
    \item Task 2: Which cluster has the largest portion overlapped by cluster \textit{C}, \textit{A} or \textit{B}?
    \item Task 3: Which cluster overlaps with the largest number of clusters?
\end{itemize}

\paragraph{Measurements} For each test, we measured performance in
terms of completion time and accuracy (error rate). 

\paragraph{Procedure} 

We randomly divided participants into two groups of ten people. The first group completed the tasks with Clusterplots, while the second group completed with Matrix. Overall, each participant completed 27 trials: 3 (number of clusters) x 3 (tasks) x 3 (repeats). The order of the trials was randomized for each participant. 

%
The experiment was performed one participant at a time. Each participant was seated in a quiet room in front of a 24-inch display screen. The experiment was divided into two parts: the
introductory session and the main experiment. In the introductory part, the administrator first briefly informed the participants
about the experiment structure and collected their demographic information. Next, the administrator introduced the experiment, showed some Clusterplots and visualization of Matrices examples, and explained the key ideas and concepts such as inter-cluster relations, cluster overlap, etc. Participants were encouraged to ask any questions they may have, and were then asked to complete a series of 12 practice trials (6 for Clusterplots and 6 for Matrix), in order to let them familiarize themselves with the tasks, and to minimize possible learning 
impediments. 

The main part of the experiment included a series of trials. For each trial, the plot (either Clusterplot or Matrix Visualization) was shown at the center of the screen and the question shown below, along with multiple choice answers to the task question. Participants were instructed to select the best answers to the given questions, and to perform their tasks as quickly and accurately as possible. Completion time and correctness were recorded for each trial.

\subsection{Result}

\begin{figure}[t]
  \centering
    \includegraphics[width=1.\linewidth]{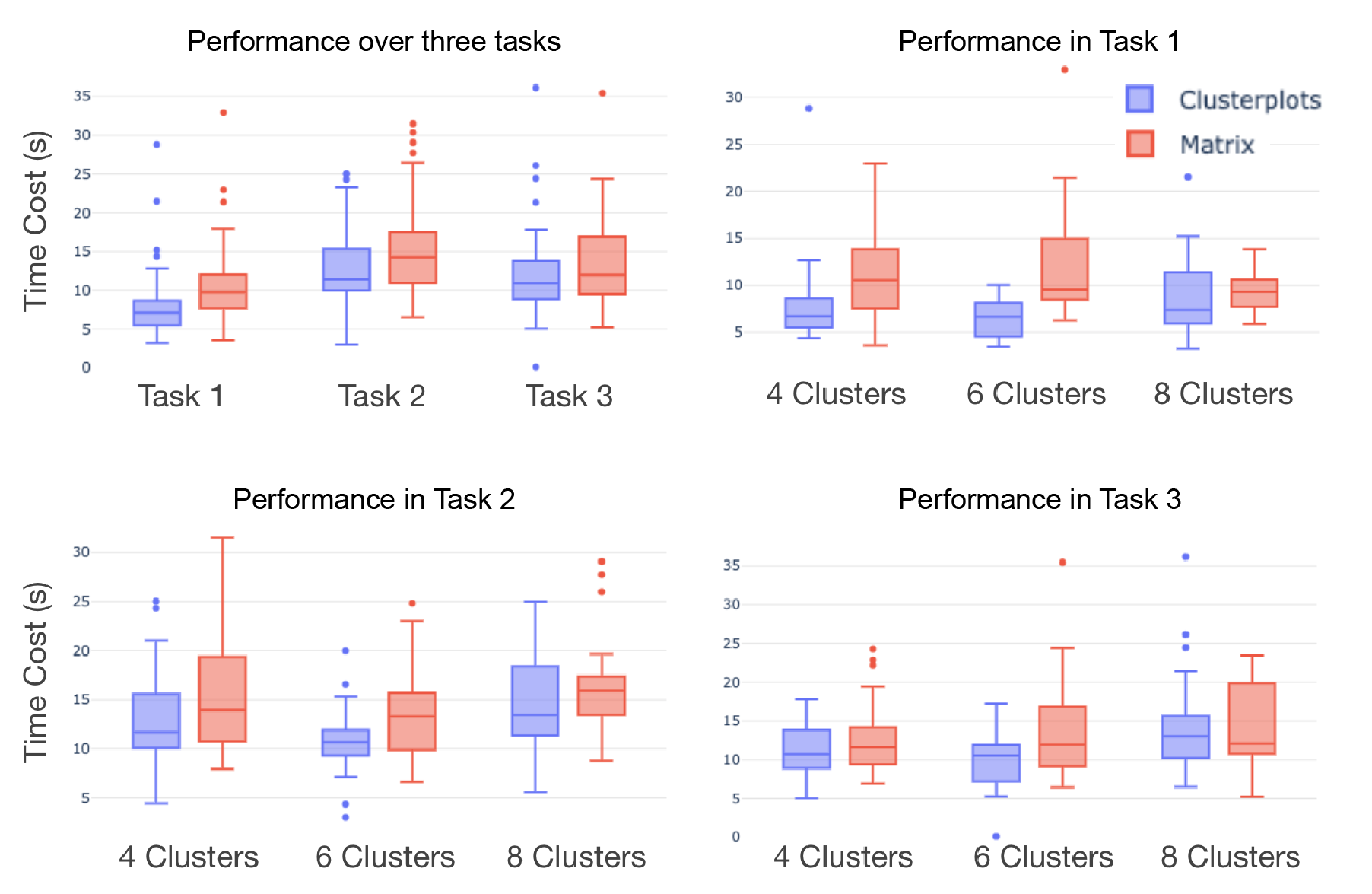}
    \caption{Time cost of two visualizations over three tasks.} 
  \label{fig:user_result}
\end{figure}

Figure~\ref{fig:user_result} summarizes the results comparing the average completion
time of Clusterplots and Matrix Visualization for the different cluster numbers over the three tasks.  
We conducted a two-way analysis of variance (ANOVA) for each task on completion time, with \textit{visualization type} (Clusterplots/Matrix) and \textit{number of clusters} (4, 6, 8) being within subject variables. As expected, a main effect was found for \textit{visualization type} that Clusterplots significantly outperforms better than Matrix over the three tasks. A main effect was found for \textit{number of cluster} in Task 2 and Task 3. Next, we report the statistical results per task separately.

\textit{Task 1} asked to identify whether two clusters overlap or not. Results indicate a strong main effect for Clusterplots, in which performance with Clusterplots (M=7.63, SD=3.62) were faster than performance using Matrix (M=10.51, SD=4.23), with F(1, 174)=24.37, p<0.001. Looking into the correctness, Task 1 is pretty simple and there were only two errors reported in Clusterplots and eight errors in Matrix. In task 1, there is no main effect reported for different cluster numbers. In \textit{Task 2}, Clusterplots (M=12.82, SD=4.55) clearly outperforms Matrix (M=15.17, SD=5.39), F(1, 174)=5.89; p=0.0016. 
\textit{Task 3} asked for a global observation of the inter-relations among clusters. Clusterplots (M=11.76, SD=5.10) averagely cost less time than Matrix (M=13.63, SD=5.44), with F(1, 174)=5.91; p=0.016. 

For the number of clusters, in \textit{Task 2}, it is found that he number of clusters effects the time cost of Clusterplots with F(2, 87)=111.09; p=0.0039, while it is not significant for Matrix visualization (F(2, 87)=1.356; p=0.263). Similar to task 2, in \textit{Task 3}, cluster number effects the performance of Clusterplots (F(2, 87)=5.91; p=0.0014) significantly, while not significant in the Matrix form (F(2, 87)=1.063; p=0.35). The different effects of cluster number on the performance of Clusterplots and Matrix might be explained by that Matrix provides more abstract operations on rows and columns, which are irrelevant to the overall number of clusters. But still, the results of evaluation show that though valuable information they may contain, these matrices do lack the ability to deliver a visual representation of the global spatial relationships among clusters. In answer, Clusterplots display the information contained in these matrices by employing a more intuitive and vivid spatial visual encoding. 



\section{Conclusions}

In this paper, we presented Clusterplots, a visualization tool for the exploration of multi-class high-dimensional datasets. Our clusterplots consists of smooth blobs that aims preserving the global geometry of the clusters and reflects the inter-class overlap and proximity, as measured in the high-dimensional data. 


It should be stressed that encoding pairwise relations in the form of a matrix can also express the relation between the clusters. However, unlike clusterplot, matrices lack spatial and geometric cues. Queries like, \emph{which large cluster has no overlap with any other cluster} or \emph{which two small clusters have significant overlap}, cannot be easily answered by simple matrix forms like for example, Confusion matrices. Queries that involve three or more clusters are even harder to be answered by tabular data. For example, \emph{which three clusters have mutual overlap} or \emph{do three specific clusters have any common region}. Moreover, clusterplot provide stronger intuition about global relation among the cluster. This global view was clearly demonstrated in Figure \ref{fig:MNIST4_AE} and Figure \ref{fig:MNISTDeepFull}(a and b), where one can immediately tell which plot represents a more confusing cluster configuration.

In summary, unlike common unsupervised techniques, which seek after the latent clusters, clusterplots use a supervised dimensionality reduction technique that uses the given class information. Clusterplot aims to provide a clear visual means to understand the nature of the inter-class relations, and to manifest the overlap and proximity between the classes. It should be noted that while the preservation of the inter-cluster proximity is inherent to all dimensionality reduction technique, the overlap preservation is where clusterplots excel as demonstrated in our results and user-study.


We have presented the effectiveness of clusterplot on several datasets, some of which are toy datasets in 3D where one can have a mental understanding of the cluster relationship and visually appreciate the expressive power of the clusterplot. We also demonstrated the competence of the clusterplot to reflect the expected performance of neural networks in generating descriptive deep features, and possibly assist Explainable AI (XAI) tools.

In the future, we would consider extending Clusterplot to include information about the density of the clusters, possibly by different transparencies or stippling. Another improvement can embed the information about the loss or uncertainty, the absolute difference between the overlap in high-dim to 2D, in the plot. For example, by changing the thickness of the border or different transparency of the blob according to the loss.

{\small
\bibliographystyle{ieee_fullname}
\bibliography{ssmct}
}

\end{document}